\documentclass[structabstract]{aa}

\usepackage{graphicx}
\usepackage{txfonts}
\usepackage{url}

\usepackage{natbib}

\bibpunct{(}{)}{;}{a}{}{,}

\begin{document}

\title{``TNOs are Cool'': A survey of the trans-Neptunian region}
\subtitle{{\bf XIV.} Size/albedo characterization of the Haumea family observed with \emph{Herschel}\thanks{{\it Herschel}
is an ESA space observatory with science instruments provided by a
European-led Principal Investigator consortia and with important participation from NASA.} and
\emph{Spitzer}}

\author{E. Vilenius\inst{1,2} \and J. Stansberry\inst{3} \and T. M\"uller\inst{2}
\and M. Mueller\inst{4,5} \and C. Kiss\inst{6} \and
P. Santos-Sanz\inst{7} \and M. Mommert\inst{8,9,12} \and
A. P\'al\inst{6} \and
E. Lellouch\inst{10} \and
J.~L. Ortiz\inst{7} \and
N. Peixinho\inst{11} \and
A. Thirouin\inst{12} \and
P. S. Lykawka\inst{13} \and
J. Horner\inst{14} \and
R. Duffard\inst{7} \and
S. Fornasier\inst{10} \and A. Delsanti\inst{15}}

\institute{Max-Planck-Institut f\"ur Sonnensystemforschung, Justus-von-Liebig-Weg 3, 37077 G{\"o}ttingen, Germany\\
\email{vilenius@mps.mpg.de}
\and
Max-Planck-Institut f\"ur extraterrestrische Physik, Postfach 1312, Giessenbachstr., 85741 Garching, Germany
\and
Space Telescope Science Institute, 3700 San Martin Drive, Baltimore, MD
21218, USA
\and
SRON Netherlands Institute for Space Research, Postbus 800, 9700 AV Groningen, the Netherlands
\and
Rijksuniversiteit Groningen, Kapteyn Astronomical Institute, Postbus 800, 9700 AV Groningen, The Netherlands
\and
Konkoly Observatory, Research Centre for Astronomy and Earth Sciences, Konkoly Thege 15-17, H-1121 Budapest, Hungary
\and
Instituto de Astrof\'isica de Andaluc\'ia (CSIC), Glorieta de la Astronom\'ia s/n, 18008-Granada, Spain
\and
Deutsches Zentrum f\"ur Luft- und Raumfahrt e.V., Institute of Planetary Research, Rutherfordstr. 2, 12489 Berlin, Germany
\and
Northern Arizona University, Department of Physics and Astronomy, PO Box 6010, Flagstaff, AZ, 86011, USA
\and
LESIA, Observatoire de Paris, Universit\'e PSL, CNRS, Univ. Paris Diderot, Sorbonne Paris Cit\'{e}, Sorbonne Universit\'e, 5 Place J. Janssen, 92195 Meudon Pricipal Cedex, France
\and
CITEUC -- Centre for Earth and Space Science Research of the University of Coimbra,
Observat\'orio Astron\'omico da Universidade de Coimbra, 3030-004 Coimbra, Portugal
\and
Lowell Observatory, 1400 W Mars Hill Rd, 86001, Flagstaff, Arizona, USA
\and
School of Interdisciplinary Social and Human Sciences, Kindai
University, Shinkamikosaka 228-3, Higashiosaka-shi, Osaka, 577-0813, Japan
\and
Centre for Astrophysics, University of Southern Queensland, Toowoomba, Queensland 4350, Australia
\and
Aix Marseille Universit{\'e}, CNRS, LAM (Laboratoire d'Astrophysique de Marseille) UMR 7326, 13388, Marseille, France
}

\date{Received 29 December, 2017 / Accepted 3 July, 2018}

\abstract
{A group of trans-Neptunian objects (TNO) are dynamically
related to the dwarf planet 136108 Haumea. Ten of them show strong
indications of water ice on their surfaces, are assumed to have
resulted from a collision, and are accepted as
the only known TNO collisional family. Nineteen other dynamically similar objects
lack water ice absorptions and are hypothesized to be dynamical interlopers.}
{We have made observations to determine sizes and
geometric albedos of six of the accepted Haumea family
members and one dynamical interloper. Ten other dynamical interlopers have been measured
by previous works. We compare the individual and
statistical properties of the family members and interlopers, examining the size and albedo
distributions of both groups. We also examine implications for the
total mass of the family and their ejection velocities.}
{We use far-infrared space-based telescopes to observe the target TNOs near their thermal peak
and combine these data with optical magnitudes to derive
sizes and albedos using radiometric techniques. Using measured and inferred sizes together with
ejection velocities we determine the power-law slope of ejection
velocity as a function of effective diameter.}
{The detected Haumea family members have a diversity of
geometric albedos $\sim$ 0.3-0.8, which are higher than
geometric albedos of dynamically similar objects without water ice.
The median geometric albedo for accepted family members is $p_V=0.48_{-0.18}^{+0.28}$,
compared to 0.08$_{-0.05}^{+0.07}$ for the dynamical interlopers.
In the size range $D=175-300$ km, the slope of the cumulative
size distribution is $q$=3.2$_{-0.4}^{+0.7}$ for accepted family members,
steeper than the $q$=2.0$\pm$0.6 slope for the dynamical interlopers with D$<$500 km.
The total mass of Haumea's moons and family members is 2.4\% of Haumea's mass.
The ejection velocities required to emplace them on their
current orbits show a dependence on diameter, with a power-law slope of 0.21-0.50.}
{}

   \keywords{Kuiper belt: general --
             Infrared: planetary systems --
             Methods: observational --
             Techniques: photometric
               }

\maketitle

\section{Introduction}
Over the past 25 years, a large number of icy bodies have been discovered orbiting beyond Neptune in the
outer solar system. These trans-Neptunian objects (TNO) are material left behind from
the formation of our solar system, and contain
a wealth of
information on how the planets migrated
to their current orbits. In addition, they likely constitute the
principal source of short-period comets,
through their daughter population, the centaurs \citep{Levison1997,Horner2004}.
The dwarf planet 136108 Haumea is one of the largest TNOs.
With a volume-equivalent diameter of $D$$\sim$1600 km
\citep{Ortiz2017},
its size is between the category of
Pluto and Eris \citep[D$>$2300 km,][]{Sicardy2011} and
the other largest TNOs 2007 OR$_{10}$, Makemake, Quaoar, and Sedna
\citep{Ortiz2012a,SantosSanz2012,BragaRibas2013,Pal2012,Pal2016}.
While mutual collisions have shaped the size distribution of small and moderate sized TNOs (diameter $<$50-100 km)
larger TNOs have generally not been eroded by disruptive collisions, so their size distribution
is thought to reflect the accretion process \citep{Davis1997}. Large objects usually experience impact cratering
instead of disruptive collisions.
However, the large object Haumea
may be an exception to this rule as it is hypothesized to be the parent body of
the so-far only identified collisional family
among TNOs \citep{Brown2007, Levison2008, Marcus2011}.
It has a short rotation period of 3.92 h \citep{Rabinowitz2006} close to the calculated and observed spin breakup limit
of TNOs \citep{Leinhardt2010,Thirouin2010}
as well as a rotationally deformed shape and a ring \citep{Ortiz2017},
which all are unique properties among the $D$$\geq$1000 km TNOs. The geometric albedo
of Haumea
($\sim$0.5) due to water ice
is
less than the
albedos of Pluto and Eris,
which have volatile ices,
whereas smaller TNOs with measured albedos available in the literature have geometric
albedos $\lesssim$0.4 \citep[e.g.][]{Lacerda2014b}.
All TNOs with $D$$\geq$1000 km for which spectra have been obtained feature methane ice on their surfaces,
except Haumea which has only water ice \citep[and references cited therein]{Barucci2011}.
Spectral modelling suggests a 1:1 mixture of crystalline and amorphous water ice on Haumea's surface
and that it
is depleted in carbon-bearing materials besides CH$_4$ compared to most other TNOs \citep{Pinilla2009}.

\citet{Brown2007} noted that a group of five TNOs including Haumea that
have
very deep near-infrared (NIR)
water ice absorption features
are also
dynamically clustered, that is, they have similar proper orbital elements.
\cite{Ragozzine2007} listed
objects with low velocities relative to Haumea's supposed collisional location.
About one third of them have strong water ice features and so are family members.
At that time
it was also known that the larger moon Hi'iaka has a strong water ice absorption
in its spectrum \citep{Barkume2006}. \citet{Brown2007} proposed that the group
of five objects are fragments of Haumea's ice mantle disrupted by a collision
with an object 60\% of the size of proto-Haumea. Such a collision may have removed
$\sim$20\% of Haumea's initial mass.
To date most authors have accepted the hypothesis
that only those TNOs which both (i) are in the dynamical cluster and (ii)
have strong water ice absorptions are members of the
family.
While some  other TNOs have water ice absorptions \citep{Brown2012b}, they are
weaker, and those TNOs are not part of the dynamical cluster.
One member of the dynamical cluster is the $D$$\sim$300 km TNO
2002 TX$_{300}$ with high geometric albedo of 0.88 \citep{Elliot2010}, which
has been identified as one of the Haumea family members as it has strong
water ice absorption bands~\citep{Licandro2006}. The whole population
of TNOs in general has a wide
range of colours (e.g. \citealp{Doressoundiram2008,Hainaut2012}) but all the
Haumea family members show neutral colours. Spectroscopic data is not available
for all potential Haumea family members and new techniques to detect water ice
signatures with NIR
photometry have been developed (e.g.
\citealp{Snodgrass2010,Trujillo2011}) in order to infer family membership. The
number of spectroscopically or photometrically confirmed members is currently
ten in addition to Haumea and its two moons Hi'iaka and Namaka
\citep{Brown2007,Ragozzine2007,Schaller2008, Snodgrass2010,Trujillo2011,Fraser2009}.

The semi-major axes of the orbits of the Haumea family members are 42.0$<$a$<$44.6 AU,
their orbital inclinations are 24.2$\degr$$<$$i$$<$29.1$\degr$, and their eccentricities are 0.11$<$$e$$<$0.17.
For all the members in the dynamical cluster the orbital elements are
40$<$a$<$47 AU, 22$\degr$$<$i$<$31$\degr$ and 0.06$<$e$\leq$0.2.
Haumea has a more eccentric orbit than the rest of the family with $e$=0.20.
It is currently in a 12:7 mean motion resonance with Neptune \citep{Lykawka2007},
and \citet{Brown2007} suggest that its current proper orbital elements have changed since
the presumed collision event. \citet{Lykawka2007} indicated that 19308 (1996 TO$_{66}$) is in a 19:11
resonance with Neptune but this resonance membership could not be confirmed by later works
\citep[e.g. ][]{Lykawka2012}.
Unless in mean motion resonance, the confirmed family members are in the dynamically hot sub-population
of classical Kuiper belt objects (CKBO) according
to the \citet{Gladman2008} classification system, but are classified as scattered-extended
in the Deep Ecliptic Survey classification system \citep{Elliot2005}.
Collisions in the present classical transneptunian belt are very unlikely
and the family would probably have been dispersed during the chaotic migration phase of planets
if it formed before the dynamically hot CKBOs had evolved to their current
orbits as predicted by the Nice model \citep[e.g.][]{Levison2008b}.
Based on calculations of collision probabilities, \citet{Levison2008}
showed that over 4.6 Ga a collision leading to the formation of one family is likely
if both the colliding objects were scattered-disk objects on highly
eccentric orbits, and that it could result in a CKBO-type orbit after the collision.

One of the biggest challenges to the collisional disruption formation mechanism is that the
objects with strong water ice features are tightly clustered, having a velocity dispersion
clearly smaller \citep[$\sim$20 $ms^{-1}$--300 $ms^{-1}$,][]{Ragozzine2007} than
the escape velocity of Haumea ($\sim$900 ms$^{-1}$).
This is unusual for fragments of a disruptive impact \citep{Schlichting2009}.
Various models have been proposed to explain the small velocity dispersion: a grazing
impact of two equal-sized objects followed by merger \citep{Leinhardt2010};
disruption of a large satellite of the proto-Haumea \citep{Schlichting2009};
and rotational fission \citep{Ortiz2012b}.
While the collisional models can explain the low velocity dispersion of the canonically-defined family
members, another possibility is that the family is more extensive than has
been assumed based on NIR spectral evidence.
A recent review of collisional mechanisms has been presented by
\citet{CampoBagatin2016}. They also propose the alternative that Haumea
together with its moons was formed independently of the family of objects
presumed to form the rest of the Haumea
family, that is, that there were two parent
bodies on close orbits.
The different water ice fractions on the surfaces of Haumea compared
to the family average found by \citet{Trujillo2011} would be compatible with this
hypothesis. The inverse correlation of size (via its proxy, the absolute magnitude)
with the presence of water ice was explained by \citet{Trujillo2011} to be
caused by two possibilities: smaller objects having a larger fraction of ice
on their surfaces or smaller objects having a larger grain size.

In order to quantify the albedos and sizes of Haumea family members
we use all available far-infrared observations.
Six of the confirmed family members have been observed with the \emph{Herschel Space
Observatory} \citep{Pilbratt2010} and four of them have also \emph{Spitzer Space Telescope}
observations. The radiometric results of five confirmed family members
19308 (1996 TO$_{66}$), 24835 (1995 SM$_{55}$), 120178 (2003 OP$_{32}$),
145453 (2005 RR$_{43}$), and 2003 UZ$_{117}$ are new in this work.
We describe these \emph{Herschel} and \emph{Spitzer} observations
as well as optical absolute magnitudes in Sect.~2 and present the radiometric analysis in
Sect.~3. We discuss the implication to the Haumea family in Sect.~4 and make conclusions
in Sect.~5.

\section{Observations and auxiliary data}

\subsection{\emph{Herschel} observations}
\label{Hobs}
The observations of the Haumea family with the \emph{Herschel Space Observatory} were part of the Open Time
Key Program ``TNOs are Cool'' \citep{Muller2009}, which used in total about 400 hours of observing time
during the Science Demonstration Phase and Routine Science Phases to observe 132 targets.
Haumea itself was observed extensively, more than ten hours with two photometric instruments,
the Photodetector Array Camera and Spectrometer (PACS) at 70, 100, and 160 $\mu m$
\citep{Poglitsch2010}
and the Spectral and Photometric Imaging Receiver (SPIRE) at
250, 350, and 500 $\mu m$ \citep{Griffin2010}.
The thermal light curve of the system of Haumea and its moons
were analysed by \citet{Lellouch2010} and \citet{SantosSanz2017} and the averaged
multi-band observations by PACS and SPIRE in \citet{Fornasier2013}.
Six confirmed Haumea family members were
observed by \emph{Herschel} as part of this work
(Table~\ref{table_obs}) using a total of about 12 hours. In addition,
eight probable dynamical interlopers\footnote{Interlopers
\citep[as defined by][]{Ragozzine2007} belong to the same dynamical cluster as Haumea
family members but they lack the spectral features to be confirmed as family
members.} were analysed in previous works from ``TNOs are Cool'' and one
of them (1999 KR$_{16}$) has updated flux densities given in
Table~\ref{table_obs}. The previously unpublished \emph{Herschel}
observations of the dynamical interloper 1999 CD$_{158}$
are part of this work.

The \emph{Herschel}/PACS observations of the Haumea family were planned in the same way as other observations
in the key programme
(e.g. \citealp{Vilenius2012}). The instrument was continuously
sampling while the telescope
moved in a pattern of parallel scan legs, each 3$\arcmin$
in length\footnote{The observations in February 2010 were done with a scan leg length of 2.5$\arcmin$.},
around the target coordinates. We had checked the astrometric uncertainty
of the coordinates with the criterion
that the 3$\sigma$ positional uncertainty was less than 10$\arcsec$.
Each PACS observation (identified by ``OBSID'') produced a map that was the result of repeating the scan pattern
several times. This repetition factor was a free parameter in the planning of the duration of observations.
In the beginning of the Routine Science Phase of \emph{Herschel} in the first half of 2010 (Table~\ref{table_obs}),
we used repetition factors of two to three based on detecting thermal emission of an object assuming it has a
geometric albedo of 0.08. Later in 2011 we used
longer observing time with repetition factors of  four to five to take into account
the possible high albedo of Haumea family members as indicated by
\citet{Elliot2010} for 2002 TX$_{300}$ because
higher geometric albedo at visible wavelengths means less emission in the
far-infrared wavelengths.

We used the Herschel Interactive Processing Environment (HIPE\footnote{Data presented in this paper were
analysed using ``HIPE'', a joint development
by the Herschel Science Ground Segment Consortium, consisting of ESA, the
NASA Herschel Science Center, and the HIFI, PACS and SPIRE consortia.}, version 9.0 / CIB 2974)
to produce Level 2 maps with the scan map pipeline script, with TNO-specific parameters given in \citet{Kiss2014}.
This script projects pixels of the original frames produced by the detector into pixels of a sub-sampled
output map.
Each target was observed with the same sequence of individual OBSIDs at two epochs separated by about one day
so that the target had moved by 25-50$\arcsec$. We applied background subtraction using the double-differential
technique \citep{Kiss2014} to produce final maps from individual OBSIDs. We used standard
aperture photometry techniques to determine flux densities.
The uncertainties were determined by implanting 200 artificial sources in the vicinity of the real source
and calculating the standard deviation of flux densities determined from these artificial sources.
The upper limits in Table~\ref{table_obs} are 1$\sigma$ noise levels of the final map determined by this
artificial source technique.
The colour corrections were calculated in the same iterative way as in \citet{Vilenius2012} and they amount to
a few percent.
The uncertainties include the absolute calibration uncertainty, which is
5\% in all PACS bands \citep{Balog2014}.

The previously published \emph{Herschel} observations of 1999 KR$_{16}$
\citep{SantosSanz2012} have been re-analysed in this work (Table~\ref{table_obs}).
\citet{SantosSanz2012} used the super-sky subtraction
method
\citep{Stansberry2008} and reported flux densities of
5.7$\pm$0.7 / 3.5$\pm$1.0 / 4.6$\pm$2.2 mJy, which were "mutually inconsistent"
as shown in their Fig. 1. In our updated analysis we found out that there was
a background source near the target located in such a way that the
double-differential
technique \citep{Kiss2014} did not fully remove it.
We consider the visit 2 images as contaminated and use only visit 1. Moreover,
we consider the 160 $\mu$m band an upper limit.

\begin{table*}
\caption{\emph{Herschel} observations and monochromatic flux densities of six unpublished and two reanalysed targets.
Targets 2002 TX$_{300}$ \citep{Lellouch2013} and 1999 KR$_{16}$ \citep{SantosSanz2012} have been reanalysed and their flux
densities updated in this work.}
\centering
\begin{tabular}{llrlcccrrr}
\hline\hline
Target & 1st OBSIDs   & Dur.  & Mid-time & $r$  & $\Delta$ & $\alpha$ & \multicolumn{3}{c}{Flux densities (mJy)} \\
       & of visit 1/2 & (min) &          & (AU) & (AU)     & (\degr)  & $70\,\mathrm{\mu m}$ & $100\,\mathrm{\mu m}$ & $160\,\mathrm{\mu m}$ \\
\hline
\object{1995 SM$_{55}$}  & 1342190925/...0994 &  73.1 & 2010-Feb-22 11:58 & 38.62 & 38.99 & 1.37 & $<$1.7        & $<$1.7        & $<$2.7 \\
\object{2005 RR$_{43}$}  & 1342190957/...1033 &  73.1 & 2010-Feb-23 00:16 & 38.73 & 38.88 & 1.45 & $2.6 \pm 1.8$ & $4.6 \pm 2.4$ & $<$2.8 \\
\object{2003 UZ$_{117}$} & 1342190961/...1037 & 109.3 & 2010-Feb-23 01:07 & 39.27 & 39.50 & 1.41 & $2.0 \pm 1.6$ & $<$2.2        & $<$2.3 \\
\object{2003 OP$_{32}$}  & 1342197669/...7721 &  75.7 & 2010-Jun-03 20:31 & 41.53 & 41.31 & 1.39 & $1.7 \pm 1.5$ & $<$2.1        & $<$4.1 \\
\object{2002 TX$_{300}$} & 1342212764/...2802 & 188.5 & 2011-Jan-17 03:46 & 41.68 & 41.76 & 1.36 & $1.2 \pm 1.1$ & $<$2.8        & $<$4.1 \\
\object{1996 TO$_{66}$}  & 1342222430/...2481 & 188.5 & 2011-Jun-10 11:25 & 46.92 & 47.34 & 1.14 & $<$1.2        & $<$1.3        & $<$2.9 \\
\hline
\object{1999 CD$_{158}$} & 1342206024/...6060 & 150.9 & 2010-Oct-08 05:01 & 47.40 & 47.83 & 1.09 & $<$1.3        & $<$1.6        & $<$2.1 \\
\object{1999 KR$_{16}$}  & 1342212814/...3071 & 188.5 & 2011-Jan-18 06:14 & 35.76 & 36.06 & 1.51 & 4.2 $\pm$1.1\tablefootmark{a} & 6.9 $\pm$2.2\tablefootmark{a} & $<$4.5 \\
\hline
\end{tabular}
\label{table_obs}
\tablefoot{OBSIDs are observation identifiers in the \emph{Herschel} Science Archive.
The first OBSID of the consecutive OBSIDs/visit are given. Duration is the total
duration of the two visits (70 $\mu$m and 100 $\mu$m filters were used for
half of the duration each), mid-time is the mean UT time, $r$ is the heliocentric
distance at mid-time, $\Delta$ is the {\it Herschel}-target distance at mid-time, and $\alpha$ is
the Sun-target-{\it Herschel} phase angle at mid-time (JPL Horizons Ephemeris System, \citealp{Giorgini1996}).
Flux densities are colour-corrected and the 1$\sigma$ uncertainties include the absolute calibration
uncertainty of 5\% in all bands.
Targets above  the horizontal line are confirmed Haumea family members while those below the line are
probable dynamical interlopers.
\tablefoottext{a}{Differential fluxes from visit 1 only. During visit 2 a background source was near the
target location. This background source is close to the edge of the images from visit 1 and could not be properly
compensated by the positive and negative images.}}
\end{table*}

\subsection{\emph{Spitzer} observations}
\label{Sobs}

Four members of the Haumea family were observed using the Multiband Imaging Photometer for Spitzer (MIPS,
\citealp{Rieke2004})
aboard the \textit{Spitzer Space Telescope}
\citep{Werner2004}. These observations utilized MIPS' chop-nod photometric mode
using the dedicated chopper mirror and spacecraft slews as nods, and the
spectral channels centred at $24\ \mathrm{\mu m}$ (effective monochromatic
wavelength: $23.68\ \mathrm{\mu m}$) and $70\ \mathrm{\mu m}$
($71.42\ \mathrm{\mu m}$). There is strong spectral overlap between the
70-micron channels of MIPS and PACS.

We reanalysed (Mueller et al., in prep.) the MIPS observations using the
methods described by \citet{Stansberry2007,Stansberry2008} and \citet{Brucker2009}, along with
recent ephemeris information.
Targets 2002 TX$_{300}$ and 2003 OP$_{32}$ were observed more than once and a background-subtraction method
was used to produce combined maps. The individual visits
were made within about two days of the first visit of the observed target.
Flux densities were
determined from the resulting mosaics using aperture photometry. Flux
uncertainties were estimated using two techniques, one using a standard
sky annulus, one using multiple sky apertures.

None of the Haumea family members were detected by \emph{Spitzer}.
Our analysis provides upper flux limits (see Table~\ref{table_Spitzerobs}).
We provide tighter limits based on new reduction of the data on the
non-detection of 2002 TX$_{300}$ than a previous analysis by
\citet{Stansberry2008}; the remaining observations have not been
published so far.

\begin{table*}
 \caption{\emph{Spitzer}/MIPS observations.}
 \centering
 \begin{tabular}{lrcccc|rc|rr}
  \hline\hline
  Target        & PID   &  Mid-time         & $r$   & $\Delta$ & $\alpha$ & \multicolumn{2}{c|}{MIPS $24\ \mathrm{\mu m}$ band} & \multicolumn{2}{c}{MIPS $70\ \mathrm{\mu m}$ band} \\
                &       &                   & (AU)  & (AU)     & (\degr)  & Dur. (min)    & F$_{24}$ (mJy) & Dur. (min)        & F$_{70}$ (mJy)     \\
  \hline
1995 SM$_{55}$  & 55    & 2006-Feb-18 16:27 & 38.93 & 39.03    & 1.47     & 16.5          & $<$0.045       & 22.4              & $<$3.75 \\
1996 TO$_{66}$  & 55    & 2004-Dec-26 10:22 & 46.40 & 46.22    & 1.23     & \ldots        & \ldots         & 44.8              & $<$4.66 \\
2002 TX$_{300}$ & 3283  & 2004-Dec-28 02:04 & 40.98 & 40.73    & 1.37     & 5.3           & $<$0.025       & 5.6               & $<$5.59 \\
2003 OP$_{32}$  & 30081 & 2006-Dec-07 00:49 & 41.19 & 41.15    & 1.41     & 57.5          & $<$0.015       & 33.6              & $<$4.80 \\
\hline
1999 KR$_{16}$  & 55    & 2006-Feb-18 05:51 & 36.73 & 36.65    & 1.56     & \ldots        & \ldots         & 44.8              & $<$2.24 \\
\hline
 \end{tabular}
\label{table_Spitzerobs}
\tablefoot{PID is the \emph{Spitzer} programme identifier.
Observing geometry (heliocentric distance $r$,
\emph{Spitzer}-target distance $\Delta$ and Sun-target-\emph{Spitzer} phase angle $\alpha$) is averaged
over the individual observations. The ``Dur.'' column gives
the total observing time (2002 TX$_{300}$ and 2003 OP$_{32}$ had more than one visit).
Targets above the horizontal line are confirmed Haumea family members and 1999 KR$_{16}$
is a probable dynamical interloper.}
\end{table*}

\subsection{Optical data}
\label{auxobs}
In the radiometric method we simultaneously fit flux densities and absolute magnitude $H_\mathrm{V}$
to the model of emitted flux and to the optical constraint, respectively (Equations \ref{model_emission}
and \ref{opt_constr} in Sect.~\ref{model}).
Generally, an accurate $H_\mathrm{V}$ affects mainly the accuracy of the
estimate of geometric albedo and has a weaker effect on the accuracy of the
diameter estimate when far-infrared data is available. However, in the case of high-albedo
objects the accuracy of the diameter estimate is affected more strongly by the
uncertainty in $H_\mathrm{V}$ than in the general case.

Due to their large distance, observations of TNOs from the ground or from near Earth
are always done at small Sun-target-observer phase angles and a linear phase
function is mostly used to derive $H_\mathrm{V}$ in the literature. Haumea and
four of the confirmed Haumea family members
(Table~\ref{table_overview_groundobs_phasestudy}) have been observed with dozens
of individual exposures at phase angles $\alpha$ in the range
0.3\degr$<$$\alpha$$<$1.5\degr \citep{Rabinowitz2007,Rabinowitz2008} and taking
into account and reducing short-term variability due to rotational light curves.
These carefully determined phase coefficients of the five objects are between
$\sim$0.01 mag/deg and $\sim$0.1 mag/deg with a weighted average of 0.066$\pm$0.024
mag/deg. The exact shape of a phase curve depends on scattering properties of
the surface and for example on porosity and granular structure \citep{Rabinowitz2008}.
A typical opposition spike at small phase angles $\alpha\lesssim$ 0.2$\degr$,
compared to extrapolating a linear phase curve, is a brightening of $\sim$0.1 mag
\citep[and references cited therein]{Belskaya2008}. Such a brightening
would mean a relative increase in the value of geometric albedo of
$\sim$10\%. However,
high-albedo objects with a phase curve slope
$\sog$0.04 mag/deg
already have an opposition surge that is too wide to allow a narrow spike
near zero phase angle \citep{Schaefer2009}. The average of good quality phase
slopes of Haumea and its family (Table~\ref{table_overview_groundobs_phasestudy})
is greater than the limit of $\sim$0.04 mag/deg and therefore we have not applied
the 0.1 mag brightening of $H_\mathrm{V}$ in this work.

The light curve due to rotation changes the optical brightness from the nominal
value between individual observations by PACS and MIPS and phasing of optical data with
the thermal observations is uncertain, therefore we quadratically add a light curve
effect to the uncertainties of $H_\mathrm{V}$ before thermal modelling as explained in \citet{Vilenius2012}.
This additional uncertainty is explicitly shown with the
uncertainty of $H_\mathrm{V}$ in Table~\ref{table_overview_groundobs_phasestudy}.

For targets lacking a phase curve study in the literature, we determine the linear
phase coefficient from combinations
of photometric-quality data points when available and/or data from the Minor Planet
Center (MPC),
which is more uncertain (see Table~\ref{table_overview_groundobs}).
Since these data have not been reduced for short-term variability due to
rotation, we have added an uncertainty to each data point in the way explained
above.
There is usually no data available at very small phase angles. An exception is
1996 TO$_{66}$, which has also data points at 0.05$\degr$ and 0.07$\degr$.
However,
these two points are well compatible with a linear trend and the phase slope of
0.20$\pm$0.12 mag/deg is higher than the $\sim$0.04 mag/deg limit. Thus, we
can assume that there is no narrow non-linear opposition spike.

The phase coefficients derived in this work are
compatible within uncertainties with the average TNO $\beta$=0.12$\pm$0.06 mag/deg of
\citet{Perna2013}, except 2003 SQ$_{317}$ which is discussed below. A more recent work
to determine linear phase coefficients
of a large sample of TNOs \citep{AlvarezCandal2016} found a median value of 0.10
mag/deg in a double distribution containing a narrow component and a
wider one with approximately half of TNOs belonging to each component of the
distribution. The maximum value reported was 1.35 mag/deg. The difference in
determining the phase coefficients in this work and in \citet{AlvarezCandal2016} is
that we represent, for each data point, the un-phased light curve contribution due to rotation
by an additional increase in the uncertainty of data points,
whereas \citet{AlvarezCandal2016} assume a flat probability distribution between
the minimum and maximum of short-term variability.
In Table \ref{table_overview_groundobs} we report phase slopes for seven targets
not included in \citet{AlvarezCandal2016}. The five targets that are included in
their work are compatible with our results within error bars, but those uncertainties
are sometimes relatively large. For 1999 KR$_{16}$ we have a flat phase curve
(0.03$\pm$0.15 mag/deg) with N=5 data points, whereas \citet{AlvarezCandal2016}
has a negative slope (-0.126$\pm$0.180 mag/deg) with N=4 data points.
Whilst their result is formally consistent with zero it includes a large
range of negative values, which is
difficult to explain based on known physical mechanisms. For 1999 OY$_3$ we have
a shallower slope with N=3 because we have rejected one outlier data
point.

The highest phase slope among our targets is 0.92$\pm$0.30 mag/deg
for 2003 SQ$_{317}$
with most of our data points from \citet{Lacerda2014},
who reported a high slope of 0.95$\pm$0.41 mag/deg. They also modelled the high-amplitude
light curve of this target and found that it is either a close binary or has a very elongated shape.
It should be noted that the six data points used for 2003 SQ$_{317}$ are limited to phase
angles 0.6-1.0 deg. If data for lower phase angles become available in the
future, it might change the current slope estimate.

For the candidate Haumea family members (membership
neither confirmed nor rejected) we use mostly
non-photometric quality data from the
Minor Planet Center due to the poor availability of high-quality optical data.
The light curve amplitudes are sparsely known and V-R colours are not known
for these candidate family members. When the light curve amplitude
is unknown we assume it to be 0.2 mag based on the finding of \citet{Duffard2009} that
70\% of TNOs have an amplitude less than this value.
We try to fit a phase curve slope but in four cases the result is not plausible,
or
not
reliable due to limited phase angle coverage. For those cases
we use an assumed value for the phase coefficient of $\beta$=0.12$\pm$0.06 mag/deg \citep{Perna2013}.
Given the $H_V$ uncertainties of these four targets, using this average value
instead of the average of confirmed Haumea family members from Table
\ref{table_overview_groundobs_phasestudy} would have only a minor effect on the
derived absolute magnitudes.

\begin{table*}
\centering
\caption{Absolute magnitudes from detailed
phase curve studies as well as light curve properties.}
\begin{tabular}{llllll}
\hline
\hline
Target                  & Amplitude                  & Period           & Single/double   &  $H_{\mathrm{V}}$ \tablefootmark{4,5} & Phase coefficient\tablefootmark{4} \\
                        & (mag)                      & ($\mathrm{h}$)   & peaked          &  (mag)                                & mag/\degr \\
\hline
136108 Haumea           & 0.320$\pm$0.006\tablefootmark{9} & 3.9154$\pm$0.0002\tablefootmark{8} & double\tablefootmark{8} & 0.428$\pm$0.011\tablefootmark{8} & 0.097$\pm$0.007 \\
24835 (1995 SM$_{55}$)  & 0.04$\pm$0.02\tablefootmark{7}      & 8.08$\pm$0.03\tablefootmark{1}         & double\tablefootmark{7} & 4.490$\pm$0.030$\pm$0.018            & 0.060$\pm$0.027 \\
55636 (2002 TX$_{300}$) & 0.05$\pm$0.01\tablefootmark{2}      & 8.15\tablefootmark{2}                  & double\tablefootmark{2} & 3.365$\pm$0.044$\pm$0.022            & 0.076$\pm$0.029 \\
120178 (2003 OP$_{32}$) & 0.14$\pm$0.02\tablefootmark{7} & 4.85\tablefootmark{6}  & single\tablefootmark{7} & 4.097$\pm$0.033$\pm$0.062            & 0.040$\pm$0.022 \\
145453 (2005 RR$_{43}$) & 0.06$\pm$0.01\tablefootmark{3} & 7.87\tablefootmark{3}  & single\tablefootmark{3} & 4.125$\pm$0.071$\pm$0.026            & 0.010$\pm$0.016 \\
\hline
\multicolumn{5}{r}{Average} & 0.066$\pm$0.024 \\
\hline
\end{tabular}
\label{table_overview_groundobs_phasestudy}
\tablefoot{Light curve amplitude is the peak-to-valley amplitude,
which is
taken into account in the error bars of the absolute V-band magnitude $H_{\mathrm{V}}$ from literature
when $H_{\mathrm{V}}$ is used as input in the radiometric analysis (see text).}
\tablebib{
(1)~\citet{Sheppard2003}; (2) \citet{Thirouin2012};
(3) \citet{Thirouin2010}; (4) \citet{Rabinowitz2008};
(5) \citet{Rabinowitz2007};
(6) \citet{Benecchi2013}; (7) \citet{Thirouin2016};
(8) \citet{Rabinowitz2006};
(9) \citet{Lockwood2014}.}
\end{table*}

\begin{table*}
\centering
\caption{Absolute magnitude based on a linear phase curve fit derived in this work.}
\begin{tabular}{lcccccccl}
\hline
\hline
Target          & V              & R                   &  N & Phase coeff.                   & $\chi^2_r$ & L.c. $\Delta m_\mathrm{R}$                              & $H_{\mathrm{V}}$  & V-R \\
                & ref.           & ref.                &    & (mag/\degr)                    &  & (mag)                                                             & (mag)             & (mag)  \\
\hline
\multicolumn{9}{c}{Confirmed family members} \\
\hline
1996 TO$_{66}$  & 5,7--8,13--16,18 & 6,17                & 9  & 0.20$\pm$0.12                & 1.6  & 0.26$\pm$0.03\tablefootmark{2}    & 4.81$\pm$0.08$\pm$0.11 & 0.389$\pm$0.043 \\
1999 OY$_3$     & 9-10           & 11, 24              & 3  & 0.013$\pm$0.079                  & 3.4 & 0.08\tablefootmark{26}                     & 6.61$\pm$0.07     & 0.345$\pm$0.046 \\
2005 CB$_{79}$  & \ldots         & 11-12, MPC          & 21  & 0.09$\pm$0.08\tablefootmark{b} & 0.6 & 0.05$\pm$0.02\tablefootmark{26}           & 4.67$\pm$0.07     & 0.37$\pm$0.05\tablefootmark{11} \\
2009 YE$_7$     & \ldots & 3\tablefootmark{a}, MPC     & 20  & (aver. Table 3)\tablefootmark{d} & 0.5  & 0.06$\pm$0.02\tablefootmark{26}        & 4.65$\pm$0.15     & (assumed) \\
2003 SQ$_{317}$ & \ldots         & 11,27               & 6  & 0.92$\pm$0.30\tablefootmark{b}  & 1.0 & 0.85$\pm$0.05\tablefootmark{27}  & 6.47$\pm$0.30 & (assumed) \\
2003 UZ$_{117}$ & 4,12,20--21    & \ldots              & 6  & 0.11$\pm$0.11                  & 0.3 & 0.2\tablefootmark{12}                  & 5.23$\pm$0.12$\pm$0.09 & \ldots \\
\hline
\multicolumn{9}{c}{Probable dynamical interlopers} \\
\hline
1999 CD$_{158}$ & 10, 25         & 11,25               & 4  & 0.05$\pm$0.80                   & 0.1 & 0.49$\pm$0.03\tablefootmark{26}  & 5.35$\pm$0.63$\pm$0.22 & 0.520$\pm$0.053\\
1999 KR$_{16}$  & 28             & 17,22--24           & 5  & 0.03$\pm$0.15                &  0.8    & 0.18$\pm$0.04\tablefootmark{22}           & 6.24$\pm$0.13$\pm$0.08 & 0.738$\pm$0.057 \\
\hline
\multicolumn{9}{c}{Candidate family members} \\
\hline
1998 HL$_{151}$ & \ldots         & MPC                 & 15 & 0.63$\pm$0.50\tablefootmark{b}   & 0.1 & (assumed)                                                 & 7.88$\pm$0.39     & (assumed) \\
1999 OK$_{4}$   & \ldots         & MPC                 & 8  & (assumed)\tablefootmark{c}       & 0.05 & (assumed)                                                & 7.69$\pm$0.26     & (assumed) \\
2003 HA$_{57}$  & \ldots         & MPC                 & 9  & (assumed)\tablefootmark{c}       & 0.2 & 0.31$\pm$0.03\tablefootmark{26}                     & 8.21$\pm$0.25     & (assumed) \\
1997 RX$_9$     & \ldots         & 1, MPC              & 11 & 0.22$\pm$0.31\tablefootmark{b}   & 0.2 & (assumed)                                               & 8.31$\pm$0.22     & (assumed) \\
2003 HX$_{56}$  & 11             & MPC                 & 8  & 0.41$\pm$0.61\tablefootmark{b}   & 0.2 & $>$0.4\tablefootmark{26}                                 & 7.00$\pm$0.56     & (assumed) \\
2003 QX$_{91}$  & \ldots         & MPC                 & 5  & (assumed)                        & 1.0 & (assumed)                                                & 7.87$\pm$0.67     & (assumed) \\
2000 JG$_{81}$  & 3              & MPC                 & 4  & 0.01$\pm$0.28                    & 3.9 & (assumed)                                                & 8.10$\pm$0.45     & (assumed) \\
2008 AP$_{129}$ & \ldots         & MPC                 & 13 & (assumed)                        & 0.5 & 0.12$\pm$0.02\tablefootmark{26}                     & 5.00$\pm$0.22     & (assumed)  \\
2014 FT$_{71}$  & MPC            & MPC                 & 2  & 0.54$\pm$0.56\tablefootmark{e}   & n/a & (assumed)                                                 & 4.89$\pm$0.48     & (assumed)  \\
\hline
\end{tabular}
\label{table_overview_groundobs}
\tablefoot{References to data from literature and databases are listed with $N$
the total number of individual V- or
R-band data points, the assumed phase coefficient is the average of TNO phase coefficients:
0.12$\pm$0.06 \citep{Perna2013}, $\chi^2_r$ is the reduced $\chi^2$
describing the goodness of fit of the linear phase curve,
$H_{\mathrm{V}}$ is the absolute
V-band magnitude with uncertainties taking into account
light curve (L.c.) amplitude $\Delta m_R$. The default light curve amplitude is 0.2 mag \citep{Duffard2009}.
The light curve uncertainty is added to targets that have
\emph{Herschel} data and taken into account as input $H_\mathrm{V}$ in thermal modelling.
V-R colours are from MBOSS-2~\citep{Hainaut2012}
unless otherwise indicated.
The assumed V-R colour is the average of dynamically hot CKBOs from
MBOSS-2: 0.51$\pm$0.14.
\tablefoottext{a}{Data from SLOAN's r' and g' bands
converted to V or R band.}
\tablefoottext{b}{Phase coefficient at R band.}
\tablefoottext{c}{Data inconsistent and would lead to a negative phase coefficient in a free fit.}
\tablefoottext{d}{Data limited to a narrow phase angle range and would lead
                  to an implausibly high phase coefficient in a free fit.}
\tablefoottext{e}{Phase coefficient fit using 12 w-band data points from MPC in the phase angle range 0.3$<\alpha<$1.2.}
}
\tablebib{
(MPC)~Minor Planet Center, URL:$<$http://www.minorplanetcenter.net/iau/lists/TNOs.html$>$;
(1) \citet{Gladman1998};
(2) \citet{Sheppard2003};
(3) \citet{Benecchi2013};
(4) \citet{Boehnhardt2014};
(5) \citet{JewittLuu1998};
(6) \citet{Sheppard2010};
(7) \citet{Davies2000};
(8) \citet{GilHutton2001};
(9) \citet{Tegler2000};
(10) \citet{Doressoundiram2002};
(11) \citet{Snodgrass2010};
(12) \citet{Carry2012};
(13) \citet{RomanishinTegler1999};
(14) \citet{Doressoundiram2005};
(15) \citet{Barucci1999};
(16) \citet{Hainaut2000};
(17) \citet{Jewitt2001};
(18) \citet{Boehnhardt2001};
(20) \citet{deMeo2009},
(21) \citet{Perna2010};
(22) \citet{Sheppard2002};
(23) \citet{Trujillo2002};
(24) \citet{Boehnhardt2002};
(25) \citet{Delsanti2001};
(26) \citet{Thirouin2016};
(27) \citet{Lacerda2014};
(28) \citet{AlvarezCandal2016}.}
\end{table*}

\section{Analysis}
\subsection{Thermal modelling}
\label{model}
We use the same thermal model approach as in previous sample papers from the ``TNOs are Cool''
\emph{Herschel} programme (see e.g. \citealp{Mommert2012,Vilenius2014}), which is based on the
near-Earth asteroid thermal model (NEATM, \citealp{Harris1998}).
We assume that the objects
are airless and spherical in shape.
Using the few data points at far-infrared wavelenghts, as well as $H_\mathrm{V}$ we solve
for size, geometric albedo $p_\mathrm{V}$, and beaming factor $\eta$ in the equations
\begin{equation}
F(\lambda,r,\Delta,\alpha) = \frac{\epsilon\left(\lambda\right)}{\Delta^2}
\int_{S} B \left( \lambda, T\left( S, q p_\mathrm{V}, \eta, r, \alpha \right)\right) \;d\textbf{S} \cdot \textbf{u}
\label{model_emission}
\end{equation}
\begin{equation}
H_\mathrm{V} = m_\mathrm{\sun}+5 \log \left(\sqrt{\pi} a \right)-\frac{5}{2} \log \left(p_\mathrm{V} S_\mathrm{proj} \right),
\label{opt_constr}
\end{equation}
where $\lambda$ is the reference wavelength of each of the PACS or MIPS bands, $r,\Delta,\alpha$ give the observing
geometry at PACS or MIPS observing epoch (heliocentric distance, observer-target distance, and
Sun-target-observer phase angle, respectively), Planck's radiation law $B$ is integrated over the illuminated part
of the surface of the object,
$\textbf{u}$ is the unit directional vector towards the observer from the surface element $d\textbf{S}$,
$q$ is the phase integral, $p_\mathrm{V}$ is the geometric albedo, $\eta$ is the beaming factor,
and spectral emissivity is assumed to be constant $\epsilon$=0.9.
In the optical constraint Eq.~(\ref{opt_constr}) $m_\mathrm{\sun}$ is the
apparent solar magnitude at V-band (-26.76$\pm$0.02 mag,
\citealt{Bessell1998,Hayes1985}) and $a$ is the distance of one astronomical unit.
In NEATM
the non-illuminated part of the object does not contribute any flux and the
temperature distribution at points on the illuminated side is $T \left( \omega \right)$=$T_\mathrm{S}\cos^{1/4} \omega$,
where $\omega$ is the angular distance from the sub-solar point and $T_\mathrm{S}$ is the
temperature at the sub-solar point,
\begin{equation}
\label{Ts}
T_{\mathrm{S}} = \left( \frac{ \left(1-qp_\mathrm{V} \right)S_{\sun}}{\epsilon \eta \sigma r^2} \right)^{\frac{1}{4}}.
\end{equation}
Here $S_{\sun}$ is the solar constant and $\sigma$ is the Stefan-Boltzmann constant.
For the phase integral we use an empirical, albedo-dependent relation,
$q=0.336 p_V+0.479$, derived from observations of icy moons of giant planets
\citep{Brucker2009}. It can be noted that the two fitted parameters in
this relation change when new data become available. \citet{Brucker2009}
excluded Phoebe and Europa as outliers. After adding Triton \citep{Hillier1990},
Pluto, and Charon \citep{Buratti2017}, there are still two outliers in the data
set: Phoebe and Pluto. Consequently, the fitted slope would be steeper. Nevertheless,
we use the \citet{Brucker2009} formula to be consistent with previously
published results from the "TNOs are Cool"
programme.

Some objects may not be compatible with the NEATM assumption of spherical shape. If we have enough
information to assume pole orientation and shape, that is, a/b and a/c, where a, b,
and c are the semi-axes of an ellipsoid (a$>$b$>$c),
then we can calculate the integral in Eq.~(\ref{model_emission}) over the ellipsoid instead of a sphere.
The computational details of using ellipsoidal geometry in asteroid thermal models have been presented
in literature, for example by \citet{Brown1985}.

\begin{figure*}
   \includegraphics[width=17cm]{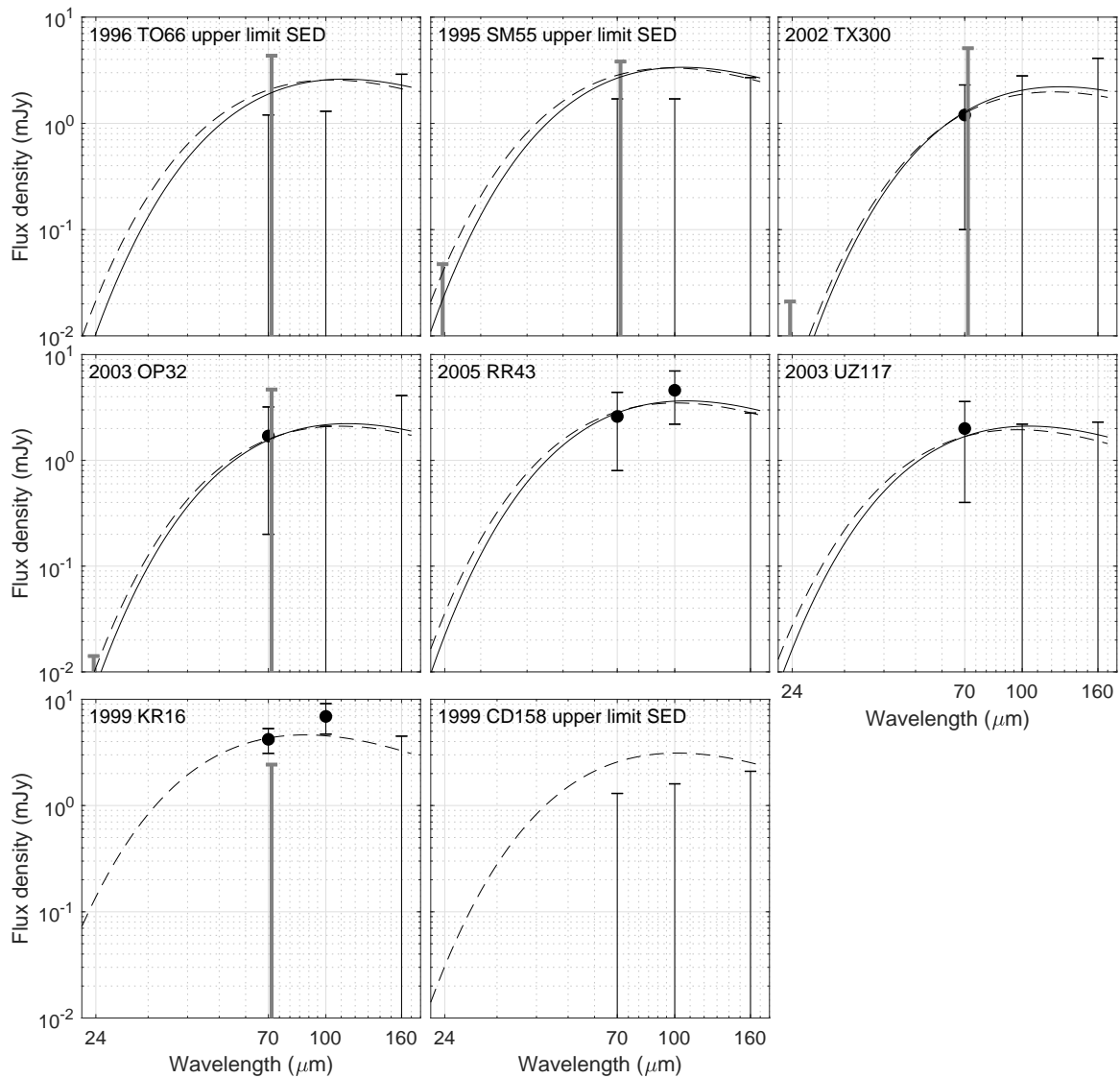}

   \caption{Modelled flux densities as function of wavelength calculated from solutions in Table~\ref{table_results}.
   Solid lines (when present) are the preferred solutions, dashed lines are fixed beaming factor solutions
   with $\eta$=1.20, except for 2002 TX$_{300}$, where the solid line has $\eta$=1.8 and the dashed line $\eta$=0.73.
   Black data points are PACS data (70, 100 and $160\ \mathrm{\mu m}$) and grey points are from MIPS
   (23.68 and $71.42\ \mathrm{\mu m}$) normalized to the observing geometry of PACS.
   Error bars are 1$\sigma$ uncertainties or 1$\sigma$ upper limits. Upper limit solutions
   have been calculated for non-detected targets using the 2$\sigma$ flux density upper limit
   of the most limiting band (see text).}
   \label{fits_lot1}
\end{figure*}

We aim to solve area-equivalent effective diameter assuming a spherical shape ($D$), $p_V$, and $\eta$ in
Eqs.~(\ref{model_emission}-\ref{opt_constr}) in the weighted least-squares parameter estimation sense,
where the weights are the squared inverses of the error bars of the measured
data points. Upper limits are replaced by a distribution
by assigning them values from a half-Gaussian distribution in a Monte Carlo
way using a set of 1000 flux density values. This technique was adopted for
faint TNOs by \cite{Vilenius2014}. The assumptions of this treatment of upper limits are that there is at least
one IR band where the target was detected and that the upper limits have a similar planned signal-to-noise ratio
as the detected band or bands. This was not the case in the PACS 160 $\mu m$ band for those
targets that were not detected in near-simultaneous PACS 100 $\mu m$ observations either.
Therefore, the 160 $\mu m$ upper limit is used only in the cases of 2005 RR$_{43}$ and 1999 KR$_{16}$.
In the other cases, where this wavelength is ignored, the solution is below
the 1$\sigma$ upper limit at 160 $\mu m$.
All the \emph{Spitzer}/MIPS flux densities are upper limits (except Haumea itself), and the
MIPS 70 $\mu m$ band observations of confirmed Haumea family members with shorter observation durations
than with the more sensitive PACS instrument have been
excluded. The MIPS 24 $\mu m$ upper limit has been included only in the modelling of 2003 OP$_{32}$ although the
solution using only PACS bands is very similar to that including also the MIPS 24 $\mu m$ upper limit.
The data sets did not allow
us to determine beaming factors and therefore we used a fixed value for $\eta$ (see Sect.~\ref{fixed_fits}).
An exception is 2002 TX$_{300}$, whose size has been measured via an occultation.
This target is discussed in Sect.~\ref{TX}. The results of radiometric fits
are given in Table~\ref{table_results}, where the last column indicates which
bands were included in the analysis of the reported solutions, which are
shown in Fig.~\ref{fits_lot1}.
Non-detected targets have been analysed in the same way as in \citet{Vilenius2014}:
the 2$\sigma$ flux limit of the most limiting band is used to derive an upper limit for effective diameter
(lower limit for geometric albedo).
For uncertainty estimates we use the Monte Carlo method of \citet{Mueller2011}
with 1000 randomized input flux densities and randomized absolute visual magnitudes as well as
randomized beaming factors in the case of fixed-$\eta$ solutions.

\begin{table*}
\centering
\caption{Results of radiometric modelling. When different fixed beaming factors have been used, the preferred solution is given in bold.}
\begin{tabular}{llcccccc}
\hline\hline
Target                           & Instruments & No. of & $D$               & $p_\mathrm{V}$  & $\eta$          & Solution         & Bands included  \\
                                 &             & bands  & (km)              &                 &                 & type             & in solution     \\
\hline
\hline
\textbf{Haumea\tablefootmark{1}} & \textbf{PACS, MIPS}  & \textbf{7}  & \textbf{2322x1704x} & \textbf{0.51$\pm$0.02} & \textbf{1.74$_{-0.17}^{+0.18}$} & \textbf{fixed $D$, $p_V$} & \textbf{all PACS, MIPS-70} \\
                                 & \textbf{SPIRE}       &             & \textbf{1026}       &                                 &                                 &               & \textbf{all SPIRE}  \\
\hline
\textbf{1996 TO$_{66}$}          & \textbf{PACS, MIPS}  & \textbf{4}        & $<$\textbf{330} & $>$\textbf{0.20} & \textbf{1.74} & \textbf{fixed $\eta$}     & \textbf{PACS-100} \\
\noalign{\smallskip}
1996 TO$_{66}$                   & PACS, MIPS  & 4      & $<$290            & $>$0.27         & $1.20$ & fixed $\eta$     & PACS-100 \\
\noalign{\smallskip}
\textbf{1995 SM$_{55}$}          & \textbf{PACS, MIPS}  & \textbf{5}      & $<$\textbf{280}   & $>$\textbf{0.36}         & \textbf{1.74} & \textbf{fixed $\eta$}     & \textbf{PACS-100} \\
\noalign{\smallskip}
1995 SM$_{55}$                   & PACS, MIPS  & 5      & $<$250            & $>$0.45         & $1.20$ & fixed $\eta$     & PACS-100 \\
\noalign{\smallskip}
\textbf{2002 TX$_{300}$}\tablefootmark{2} & \textbf{PACS, MIPS}  & \textbf{5} & \textbf{323}$_{-37}^{+95}$ & \textbf{0.76}$_{-0.45}^{+0.18}$ & \textbf{1.8}$_{-0.9}^{+0.5}$  & \textbf{fixed $D$, $p_V$} & \textbf{PACS-70, PACS-100} \\
\noalign{\smallskip}
\textbf{2003 OP$_{32}$}          & \textbf{PACS, MIPS}  & \textbf{5}      & \textbf{274}$_{-25}^{+47}$ & \textbf{0.54}$_{-0.15}^{+0.11}$ & \textbf{1.74}$\pm$\textbf{0.17} & \textbf{fixed $\eta$}     & \textbf{MIPS-24, PACS-70, PACS-100} \\
\noalign{\smallskip}
2003 OP$_{32}$                   & PACS, MIPS  & 5      & 248$_{-23}^{+32}$ & 0.66$_{-0.14}^{+0.15}$ & $1.20 \pm 0.35$ & fixed $\eta$     & MIPS-24, PACS-70, PACS-100 \\
\noalign{\smallskip}
\textbf{2005 RR$_{43}$}          & \textbf{PACS} & \textbf{3} & \textbf{300}$_{-34}^{+43}$ & \textbf{0.44}$_{-0.10}^{+0.12}$ & \textbf{1.74} $\pm$ \textbf{0.17} & \textbf{fixed} $\eta$ & \textbf{all PACS} \\
\noalign{\smallskip}
2005 RR$_{43}$                   & PACS        & 3      & 268$_{-26}^{+42}$ & 0.55$_{-0.15}^{+0.13}$ & $1.20 \pm 0.35$ & fixed $\eta$ & all PACS \\
\noalign{\smallskip}
\textbf{2003 UZ$_{117}$}         & \textbf{PACS} & \textbf{3} & \textbf{222}$_{-42}^{+57}$ & \textbf{0.29}$_{-0.11}^{+0.16}$ & \textbf{1.74} $\pm$ \textbf{0.17} & \textbf{fixed $\eta$} & \textbf{PACS-70, PACS-100} \\
\noalign{\smallskip}
2003 UZ$_{117}$                  & PACS        & 3      & 192$_{-28}^{+54}$ & 0.39$_{-0.15}^{+0.16}$ & $1.20 \pm 0.35$ & fixed $\eta$     & PACS-70, PACS-100 \\
\hline
\noalign{\smallskip}
\textbf{1999 KR$_{16}$}          & \textbf{PACS, MIPS}  & \textbf{4}      & \textbf{232}$_{-36}^{+34}$ & \textbf{0.105}$_{-0.027}^{+0.049}$         & \textbf{1.20} $\pm$ \textbf{0.35}  & \textbf{fixed $\eta$}     & \textbf{MIPS-70, all PACS} \\
\noalign{\smallskip}
\noalign{\smallskip}
\textbf{1999 CD$_{158}$}         & \textbf{PACS} & \textbf{3} & \textbf{$<$310} & \textbf{$>$0.13}              & \textbf{1.20} $\pm$ \textbf{0.35}  & \textbf{fixed $\eta$}     & \textbf{PACS-70} \\
\hline
\end{tabular}
\label{table_results}
\tablebib{(1) Size and geometric albedo from \citet{Ortiz2017}; (2) Diameter from re-analysis of the occultation result of \citet{Elliot2010} (see text).}
\end{table*}

\subsection{Haumea}
\label{haumea_eta}
The optical light curve of Haumea has a large amplitude
\citep{Rabinowitz2006},
which is indicative of a shape effect.
Time-resolved photometry shows a lower-albedo region on its surface, which may cover
more than 20\% of the instantaneous projected surface area \citep{Lacerda2008}.
\citet{Lockwood2014} observed the optical light curve by Hubble and were able
to resolve the contribution of the primary component excluding the
contribution of Haumea's moons. They report a light curve amplitude of
0.320$\pm$0.006 mag (valley-to-peak).
Using this light curve \citet{Lockwood2014} derived Haumea's size
assuming hydrostatic equilibrium, an equator-on viewing geometry, and Hapke's
reflectance model (with parameters derived for the icy moon Ariel):
$a=960$ km, $b=770$ km, and $c=495$ km for the semi-axes,
respectively. Most recently, the shape of Haumea was derived in a more direct way
from a stellar occultation \citep{Ortiz2017}: $a$=1161$\pm$30 km,
$b$=852$\pm$4 km, $c$=513$\pm$16 km. Furthermore, the new density
estimate based on this occultation result indicates that the assumption of
hydrostatic equilibrium does not apply in the case of Haumea \citep{Ortiz2017}.
The equivalent mean diameter of the projected surface corresponding to
the above mentioned ellipsoid is $2a^{1/4}b^{1/4}c^{1/2}$=
1429$\pm$22 km, which is within the uncertainty of the less accurate
radiometric spherical-shape size estimate of 1324$\pm$167 km \citep{Lellouch2010}.
However, a size estimate done by a similar method but using more data points in
far-infrared wavelengths gave a significantly smaller size of $1240_{-58}^{+69}$ km
\citep{Fornasier2013}. The geometric albedo of Haumea based on the occultation
is $p_V=0.51\pm0.02$ \citep{Ortiz2017}. Since the calculation of the geometric
albedo requires the absolute magnitude $H_V$ , \citet{Ortiz2017} used an updated
value of $H_V$ for the time of the occultation and assumed a brightness
contribution of 11\% from the two moons and 2.5\% from the ring.

In our further analysis we will use Haumea's beaming factor.
It has different values reported in the literature:
(i) 1.38$\pm$0.71 \citep{Lellouch2010} based on averaged PACS light curve data combined with a
\emph{Spitzer} observation using a NEATM-type radiometric model,
(ii) 0.95$_{-0.26}^{+0.33}$ \citep{Fornasier2013} based on a NEATM-type model and averaged data from
\emph{Herschel}/PACS as well as observations from \emph{Herschel}/SPIRE and
\emph{Spitzer}/MIPS covering a wavelength range from 70 to 350 $\mu m$, and
(iii) $\eta=0.89_{-0.07}^{+0.08}$ based on the
\citet{Lockwood2014} shape mentioned above and
\emph{Spitzer}/MIPS 70 $\mu m$ light curve using
another thermal model with isothermal temperature at each latitude
(as applied by \citealt{Stansberry2008}) as Haumea is rotating relatively quickly.
Because of differences in the radiometric models
applied, caution should be taken when comparing the beaming factor of \citet{Lockwood2014}
with the other beaming factors. \citet{Lellouch2010} modelled also the PACS light
curve of Haumea and determined the beaming factor depending
on the assumed pole orientation such that $\eta$=1.15 if Haumea is equator-on and
$\eta$=1.35 if the equator is at an angle of 15$^\circ$.

In this work, we have determined the beaming factor $\eta$ by fixing the
semi-axis and geometric albedo using the occultation result and then applying an
``ellipsoidal-NEATM'' with zero sun-target-observer phase angle
\citep{Brown1985} and far-infrared fluxes of \cite{Fornasier2013} with minor updates.
Since the measured fluxes have been obtained by averaging a
light curve or by combining at least two separate observations taken several
hours apart, we use an average projected size at
a rotation of 45$\degr$ (in a coordinate system where
rotation=0$\degr$ means that the longest axis is towards the observer).
A one-parameter fit with the ellipsoidal thermal model
gives $\eta$=1.74$_{-0.17}^{+0.18}$.
This beaming factor is higher than previous estimates when the accurate size
was not available. While Haumea's beaming factor is not unusual for objects at
$\sim$50 AU distance from the Sun, there is an observational result that other
high-albedo objects \citep[$p_V>$0.20, see Fig. 2 in][]{Lellouch2013} have
lower beaming factors with the exception of Makemake, whose beaming
factor is $\eta=2.29_{-0.40}^{+0.46}$ \citep{Lellouch2013} based on
\emph{Herschel}/SPIRE data and fixed size and geometric albedo
($p_V\approx0.77$) from a stellar occultation \citep{Ortiz2012a}. A fast rotation
tends to increase the beaming factor $\eta$ but there are also other effects
affecting $\eta$ such as increasing surface porosity, which lowers its value
\citep{Spencer1989}. With $P$=7.7 h \citep{Thirouin2010} Makemake is a slower
rotator than Haumea.

The beaming factor $\eta$ is related to the thermal parameter
$\Theta$ of \citet{Spencer1989}, which is the ratio of two characteristic timescales: the timescale of radiating heat from the subsurface and the diurnal
timescale. Figure 5 in \citet{Lellouch2013} shows the beaming factor as a
function of the thermal parameter for a spherical object with an instantaneous
subsolar temperature of $T_0=50$ K, which is close to the $T_0$ of Haumea that
can be calculated via our Eq.~\ref{Ts} by setting $\eta=1$. Furthermore, Fig. 4
of \citet{Lellouch2013} shows that the relation between the beaming factor and
the thermal parameter does not depend on small differences in the value of $T_0$
if the thermal parameter is $\Theta\lesssim$10. However, there is a strong
dependence on the aspect angle of the rotation axis and based on the occultation
Haumea is seen close to equator-on \citep{Ortiz2017}. The beaming factor
derived in this work for Haumea implies a thermal parameter $\Theta$
in the order of magnitude of $\sim$3
if there is no surface roughness and up
to a factor of approximately two
higher in case of high roughness.
Thermal inertia $\Gamma$ is directly proportional to the thermal parameter
\citep{Spencer1989}
\begin{equation}
\Gamma = \Theta \frac{\epsilon \sigma T_0^3}{\sqrt{2 \pi}} \sqrt{P},
\label{thermal_inertia}
\end{equation}
where $P$ is the rotation period given in
Table~\ref{table_overview_groundobs_phasestudy}.
This estimate gives a thermal inertia of $\Gamma\sim$1 $\mathrm{J} \mathrm{m}^{-2} \mathrm{K}^{-1}
\mathrm{s}^{-\frac{1}{2}}$, which is compatible with the finding of
\citet{Lellouch2013} that most high-albedo objects
have very low thermal inertias.\footnote{The average thermal inertia of TNOs and
centaurs, without restricting geometric albedo, is
(2.5$\pm$0.5) $\mathrm{J} \mathrm{m}^{-2} \mathrm{K}^{-1} \mathrm{s}^{-\frac{1}{2}}$ and the thermal inertia
decreases to $\sim$0.5 $\mathrm{J} \mathrm{m}^{-2} \mathrm{K}^{-1} \mathrm{s}^{-\frac{1}{2}}$ for high-albedo
objects \citep{Lellouch2013}.} The value derived in this work is higher than
the thermophysical modelling of \citet{SantosSanz2017}, which indicates
that Haumea's thermal inertia is $<$0.5 $\mathrm{J} \mathrm{m}^{-2} \mathrm{K}^{-1} \mathrm{s}^{-\frac{1}{2}}$ and
probably as low as $<$0.2 $\mathrm{J} \mathrm{m}^{-2} \mathrm{K}^{-1} \mathrm{s}^{-\frac{1}{2}}$.
\citet{SantosSanz2017} used thermal light curves observed by \emph{Herschel} as
well as the shape model and geometric albedo estimate available before the
results from the occultation were analysed. Sophisticated thermophysical
modelling using the occultation size and shape as well as contributions from
the moons, the ring, and a dark spot on Haumea is beyond this work and will be
analysed separately (M{\"u}ller et al., {\it in prep.}).
The observational
result of a lack of high beaming factors of high-albedo objects mentioned
earlier is reflected also onto thermal inertias inferred from measured
beaming factors and rotational periods: high values of thermal
inertia are excluded for high-albedo objects \citep[see Fig. 7 in][]{Lellouch2013}. In
addition to Haumea, another moderate to high-albedo TNO that has a value of
thermal inertia determined via thermophysical modelling is Orcus. Using \emph{Herschel}
observations its thermal inertia has been determined to be
0.4$<$$\Gamma$$<$2.0 $\mathrm{J} \mathrm{m}^{-2} \mathrm{K}^{-1} \mathrm{s}^{-\frac{1}{2}}$ \citep{Lellouch2013}.
Orcus has a geometric albedo of $p_V\approx$0.23 and a beaming factor of
$0.97_{-0.02}^{+0.05}$ \citep{Fornasier2013}. Haumea's thermal inertia
estimated in this work is compatible with the thermal inertia determined for
Orcus although its beaming factor is lower than
that of Haumea. With its light curve period of $\sim$10 h
\citep{Thirouin2010}, Orcus is a much slower rotator than Haumea but this difference
is not enough to explain the difference in beaming factors. Orcus is likely
to have a surface with more roughness than that of Haumea.

\subsection{Occultation target 2002 TX$_{300}$}
\label{TX}
Target 2002 TX$_{300}$ was
observed both by \emph{Herschel} and \emph{Spitzer}, but only the PACS/70 $\mu$m band gives a weak
detection while the other four bands give upper limits. Although \citet{Lellouch2013} reported
a three-band detection (all having signal-to-noise ratio $<$3), after an updated data reduction
the PACS 100 and 160 $\mu$m bands are now considered upper limits.
The \emph{Spitzer} observations of 2002 TX$_{300}$ were
of very short duration (Table~\ref{table_Spitzerobs}) compared to the \emph{Herschel} observations.
We have ignored the \emph{Spitzer} and PACS/160 $\mu$m data because those upper limits
do not constrain the solution. A floating-$\eta$ solution that would be compatible with
the optical constraint (Eq.~\ref{opt_constr}) is not possible in the physical range of the
beaming factor: 0.6$\leq$$\eta$$\leq$2.65 (limits discussed in \citealp{Mommert2012} and \citealp{Lellouch2013}).
However, for this target there is an independent size estimate available from a stellar occultation
event in 2009.

The observations of the occultation event of 2002 TX$_{300}$ by several stations resulted
in two useful chords. The diameter assuming a circular fit is 286$\pm$10 km
\citep{Elliot2010}. While the occultation technique may give very accurate
sizes of TNOs, it should be noted that in the case of 2002 TX$_{300}$ the
result is based on two chords as a reliable elliptical shape fit would require
at least three chords. In addition, the mid-times of the occultations reported by the
observing stations at Haleakala and Mauna Kea differ by 31.056 s
\citep[Table 1 in][]{Elliot2010}. Such an offset, if real, would be compatible
with a hypothesis that the two chords are from two separate objects, that is, that
2002 TX$_{300}$ could be a binary. \citet{Elliot2010} mention that one of the
chords had to be shifted by 32.95 s to get them aligned for a circular fit (fit
parameters were radius, centre
position in the sky plane relative to the occulted star, and timing offset).
The two-chord occultation and a large timing uncertainty imply a larger uncertainty also
in the adopted effective size estimate. The actual shape of an object the size of
2002 TX$_{300}$ may differ from a spherical one since self-gravity is not strong enough
for an icy $\lesssim$400 km object to result in a sphere-like shape.
The optical light curve is double-peaked which indicates a shape effect. If we
assume a Maclaurin spheroid with a rotation period of
8.15 hours and a uniform density of 1.0 gcm$^{-3}$ , the axial ratio a/c would
be 1.27 according to the figure of equilibrium formalism. This ratio is even larger
for lower densities. An ellipsoidal fit with a/c$\sim$1.3 would give
a major axis of 363 km, a minor axis of 289 km, and an effective diameter of 323 km,
which is 13\% more than the circular fit would give. Even larger effective
diameters would result if one of the chords is moved arbitrarily within the
timing shift.

The geometric albedo is calculated from the occultation size via absolute
magnitude $H_\mathrm{V}$. In this work (see
Table~\ref{table_overview_groundobs_phasestudy}) we use
$H_\mathrm{V}$=3.365$\pm$0.044 mag based on a phase curve study
(\citealt{Rabinowitz2008}), which is different from the $H_\mathrm{V}$ used by
\citeauthor{Elliot2010} ($\approx$3.48). Using the \citeauthor{Elliot2010} size
for a circular fit and the \citeauthor{Rabinowitz2008} absolute magnitude
results in a very high geometric albedo of 0.98$\pm$0.08. This is
higher than the geometric albedo of 0.88$\pm$0.06 reported by \citet{Elliot2010}
for a circular fit\footnote{\citet{Elliot2010} increased the upper albedo
uncertainty to take into account possible elliptical fits (based on
$\Delta m_R$=0.08 mag) so that the final geometric albedo was
0.88$_{-0.06}^{+0.15}$.} but is within their extended error bar when uncertainty
due to possible elliptical fits is taken into account. A geometric albedo of
$p_V$=0.98 would be the highest value among TNOs and similar to that of the
dwarf planet Eris (0.96$_{-0.04}^{+0.09}$, \citealt{Sicardy2011}).

In this work we adopt the above mentioned elliptical solution based on a/c=1.3
and use 323 km as the effective diameter. The lower uncertainty limit is
estimated as the difference of this size and the circular solution. The upper
uncertainty limit is challenging to estimate from the occultation data alone.
Here we use the fact that 2002 TX$_{300}$ is close to the detection limit of
\emph{Herschel} observations. If the PACS/70 $\mu$m data point is interpreted
as an upper limit then using the 2$\sigma$ flux limit as explained in
Sect.~\ref{model} and a conservative high beaming factor we get an upper
limit of effective diameter: 418 km. Thus, our new size estimate for
2002 TX$_{300}$ is 323$_{-37}^{+95}$ km and geometric albedo estimate
$p_V$=0.76$_{-0.45}^{+0.18}$. This geometric albedo is higher but within the
large uncertainty compared to Haumea's $p_V$=0.51$\pm$0.02 \citep{Ortiz2017}.

Using fixed estimates of diameter and geometric albedo in the thermal modelling
we can fit the beaming factor. The same approach was used by \citet{Lellouch2013}.
The new size and geometric albedo estimates given above result in a beaming
factor of $\eta$=1.8$_{-0.9}^{+0.5}$.
This is higher, but compatible
within error bars, compared to an earlier result by \citet{Lellouch2013}:
$\eta$=1.15$^{+0.55}_{-0.74}$, which is based on the smaller size and higher
geometric albedo reported by \citet{Elliot2010} as well as on an earlier
version of flux densities from \emph{Herschel}. For comparison, using the
same size estimate of \citet{Elliot2010} and geometric albedo of $p_V$=0.98
results in a beaming factor of $\eta$=0.73 using updated \emph{Herschel} fluxes
(see also Fig.~\ref{fits_lot1}).

\subsection{Fixed-$\eta$ fits}
\label{fixed_fits}
Fixed-$\eta$ solutions were used when floating-$\eta$ fits failed.
Most of the TNO literature has used the default value
$\eta$=1.20$\pm$0.35 \citep{Stansberry2008} based on a sample of TNOs of
various dynamical classes observed by \emph{Spitzer} where CKBOs were
under-represented. Based on a sample of 13 CKBOs observed by
\emph{Herschel} and/or \emph{Spitzer,} \citet{Vilenius2014} derived an average
of $\eta$=1.45$\pm0.46$. A larger sample of 85 objects observed by
\emph{Herschel} and \emph{Spitzer} representing various dynamical classes gave
a mean value of $\eta$=1.175$\pm$0.45 \citep{Lellouch2017}.

As mentioned in Sect.~\ref{model} the data quality did not allow a
floating-$\eta$ solution for most targets. Only Haumea and 2002 TX$_{300}$ have a
beaming factor determined but the latter was weakly detected only at one
thermal band (see Table~\ref{table_obs}) and has large error bars, which cover
most of the physically plausible range of beaming factor values.
Since the beaming factor depends on surface properties and heliocentric
distance \citep[e.g.][]{Lellouch2013}, we do not have a reliable average $\eta$
for the Haumea family. In this work we adopt the value of Haumea from the
one-parameter fit using the occultation size and albedo as explained in
Sect.~\ref{haumea_eta},
but approximate the asymmetric uncertainties with a symmetric Gaussian
distribution in further analysis: $\eta$=1.74$\pm$0.17.
We have adopted
this value in our fixed-$\eta$ fits for confirmed family members
(2003 OP$_{32}$, 2005 RR$_{43}$, 2003 UZ$_{117}$, and upper limits of
1996 TO$_{66}$ and 1995 SM$_{55}$), but show also the results based on the
canonical default value $\eta$=1.20$\pm$0.35 in Table~\ref{table_results}.
The rotational periods of 2003 OP$_{32}$, 2005 RR$_{43}$, and 2003 UZ$_{117}$
have been measured and we can estimate the value of their thermal parameters
(Eq.~\ref{thermal_inertia}) assuming a value for the thermal inertia.
Plausible values are 1.0$<$$\Gamma$$<$3.0 $\mathrm{J} \mathrm{m}^{-2} \mathrm{K}^{-1} \mathrm{s}^{-\frac{1}{2}}$ if the
thermal inertias of these three objects do not differ significantly from that
of Haumea's or the average thermal inertia (see Sect.~\ref{haumea_eta}). With
this range of thermal inertia, the thermal parameter is 2.2$<\Theta<$8.4 for the
three objects. Therefore, a beaming factor value of $\eta\approx1.74$ is possible
for these three objects.
We continue to use the default value of the beaming factor 1.20$\pm$0.35
for the two
moderate-albedo probable dynamical interlopers (1999 KR$_{16}$ and
1999 CD$_{158}$) modelled in this work, as the probable dynamical interlopers are
in a different cluster in a colour-albedo diagram
\citep[see Fig. 2 in][]{Lacerda2014b} and thus probably do not share
the surface properties of Haumea family members.

\subsection{Comparison with earlier results}
\label{resultscomp}
Four of the family members, in addition to Haumea, have been observed by \emph{Spitzer}.
Based on upper limits at two \emph{Spitzer}/MIPS bands, \citet{Brucker2009} reported 1$\sigma$ limits
for 2002 TX$_{300}$ as $D$$<$210 km and $p_V$$>$0.41. As discussed in
Sect.~\ref{TX} the size of this object, based on a stellar occultation, is larger (Table~\ref{table_results})
than the 1$\sigma$ upper limit by \emph{Spitzer}. The other family members do not
have published \emph{Spitzer} results (except Haumea). \citet{Altenhoff2004} observed 1996 TO$_{66}$ and
1995 SM$_{55}$ with the 30 m telescope of the Institute for Radio Astronomy in the Millimeter Range (IRAM)
at 1.2 mm wavelength. The non-detections gave limits \citep{Grundy2005}
1996 TO$_{66}$: $D$$<$$902$ km, $p_R$$>$0.033 and 1995 SM$_{55}$: $D$$<$704 km, $p_R$$>$0.067.
The results of this work give more constraining limits: both targets are
smaller than previous limits
and have moderate to high albedos (Table~\ref{table_results}).

\emph{Herschel} results of the probable dynamical interloper 1999 KR$_{16}$ have been published by
\citet{SantosSanz2012}. After significant flux updates
at 100 and 160 $\mu m$ (see Sect.~\ref{Hobs}) as well as a fainter
$H_V$, the size estimate is 9\% smaller (232$_{-36}^{+34}$ km compared to the
previous 254$\pm$37 km) but the two results are within each others uncertainties.
Geometric albedo is now slightly lower ($p_V$$=$0.105$_{-0.027}^{+0.049}$) than in
\citeauthor{SantosSanz2012} ($p_R$=0.204$_{-0.050}^{+0.070}$, which corresponds
to a V-band albedo of $p_V$$\approx$0.14 using the V-R colour from
Table~\ref{table_overview_groundobs}).

\section{Sample results and discussion}
\label{discussions}

\begin{table*}
\caption{Diameters and albedos of confirmed Haumea family members.}
\begin{tabular}{lllllll}
\hline\hline
Name                           & $\Delta v_{\mathrm{min}}$ & H$_2$O    & Thermal & Diameter & Geometric & Size/albedo \\
                               & (m s$^{-1}$)              & reference & data    & (km)     & albedo    & reference \\
\hline
136108 Haumea (2003 EL$_{61}$) & 323.5  & \citet{Brown2007}           & (S+H)  & 2322x1704      & 0.51$\pm$0.02    & O17 \\
                               &        &                             &        & x1026          &                        & \\
\noalign{\smallskip}
Hi'iaka                        & \ldots & \citet{Barkume2006}         & \ldots & 383$_{-113}^{+74}$ (*)       & default                   & TW \\
\noalign{\smallskip}
Namaka                         & \ldots & \citet{Fraser2009}          & \ldots & 193$_{-65}^{+48}$ (*)       & default                   & TW \\
\noalign{\smallskip}
19308 (1996 TO$_{66}$)         & 24.2   & \citet{Brown1999}           & (S+H)  & 210$_{-62}^{+40}$ (*)       & default                & TW \\
\noalign{\smallskip}
24835 (1995 SM$_{55}$)         & 149.7  & \citet{Brown2007}           & (S+H)  & 243$_{-71}^{+46}$ (*)       & default                & TW \\
\noalign{\smallskip}
55636 (2002 TX$_{300}$)        & 107.5  & \citet{Licandro2006}  & S+H    & $323_{-37}^{+95}$    & $0.76_{-0.45}^{+0.18}$ & TW, E10\tablefootmark{a} \\
\noalign{\smallskip}
86047 (1999 OY$_3$)            & 292.8  & \citet{Ragozzine2007}       & \ldots & 91$_{-27}^{+17}$ (*)        & default                & TW \\
\noalign{\smallskip}
120178 (2003 OP$_{32}$)        & 123.3  & \citet{Brown2007}           & S+H    & 274$_{-25}^{+47}$ & 0.54$_{-0.15}^{+0.11}$ & TW \\
\noalign{\smallskip}
145453 (2005 RR$_{43}$)        & 111.2  & \citet{Brown2007}           & H      & 300$_{-34}^{+43}$ & 0.44$_{-0.10}^{+0.12}$ & TW \\
\noalign{\smallskip}
308193 (2005 CB$_{79}$)        & 96.7   & \citet{Schaller2008}        & \ldots & 224$_{-48}^{+37}$ (*)       & default                & TW \\
\noalign{\smallskip}
386723 (2009 YE$_7$)           & 85\tablefootmark{b} & \citet{Trujillo2011} & \ldots & 226$_{-50}^{+40}$ (*) & default                & TW \\
\noalign{\smallskip}
2003 SQ$_{317}$                & 148.0  & \citet{Snodgrass2010}       & \ldots & 98$_{-24}^{+20}$ (*)        & default                & TW \\
\noalign{\smallskip}
2003 UZ$_{117}$                & 66.8   & \citet{Schaller2008}        & H      & 222$_{-42}^{+57}$ & 0.29$_{-0.11}^{+0.16}$ & TW \\
\hline
\end{tabular}
\tablefoot{* = inferred using geometric albedo of
$p_V$=0.48$_{-0.18}^{+0.28}$.
$\Delta v_{\mathrm{min}}$ (from \citealp{Ragozzine2007} unless otherwise indicated) is the minimum
of four possible solutions of velocity relative to the collision location's orbit in a calculation
where the information about the original orbital angles $\Omega$, $\omega$ and $M$ has been lost.
References to first detection of water ice confirming family membership, sources of
thermal data: S for \emph{Spitzer Space Telescope} and H for \emph{Herschel Space Observatory}
(The parentheses indicate that thermal data were not used in the size/albedo
solution shown in this table). O17=\citet{Ortiz2017}, TW=This work,
E10=\citep{Elliot2010}.
(a) The result of 2002 TX$_{300}$ is from an occultation event re-analysed in this work (see Sect.~\ref{TX}).
(b) This work.}
\label{members}
\end{table*}

\begin{table}
\caption{Candidate Haumea family members (membership
neither confirmed nor rejected).}
\begin{tabular}{lll}
\hline\hline
Target     & $\Delta v_{\mathrm{min}}$  & Class \\
           & (m s$^{-1}$)               & \\
\hline
1998 HL$_{151}$ & 142.5                 & CKBO \\
1999 OK$_{4}$   & 161.5                 & CKBO \\
2003 HA$_{57}$  & 214.3                 & Plutino \\
1997 RX$_9$     & 306.1                 & CKBO \\
2003 HX$_{56}$  & 363.2                 & CKBO \\
2003 QX$_{91}$  & 222.0 \tablefootmark{a} & Res 7:4\tablefootmark{b} \\
130391 (2000 JG$_{81}$)  & 235.1 \tablefootmark{a} & Res 2:1\tablefootmark{b} \\
315530 (2008 AP$_{129}$) & 107$\pm$2 \tablefootmark{c}   & CKBO \\
2014 FT$_{71}$  & 30$\pm$1 \tablefootmark{c}       & CKBO\tablefootmark{c,d} \\
\hline
\end{tabular}
\tablefoot{Minimum velocity relative to the collision locations's orbit $\Delta v_{\mathrm{min}}$ \citep{Ragozzine2007} as in Table~\ref{members}.
Dynamical class is according to the Gladman system \citep{Gladman2008}.
(a) $\Delta v_{\mathrm{min}}$ calculated using adjusted proper elements while conserving the proper
Tisserand parameter \citep{Ragozzine2007}. (b) Information about resonant orbits from \citet{Volk2011}.
(c) This work. (d) Influenced by 7:4 mean motion resonance.}
\label{potentials}
\end{table}

Thirty-five TNOs were identified by \cite{Ragozzine2007} as potential Haumea family members based
on their orbital dynamics and velocities
with respect to the centre of mass of the collision, which is approximated by the orbit of Haumea before diffusion
under the influence of the 12:7 mean-motion resonance with Neptune.
Tables~\ref{members} and \ref{rejected} give the albedos and diameters of the
Haumea family members and of probable dynamical interlopers that have
measurements relevant to assessing their membership in the family.
Table~\ref{potentials} summarizes ejection velocities for dynamically similar
TNOs that lack any such data, and so are candidates for membership.
The ejection velocities in Tables~\ref{members} and \ref{potentials} may be
systematically uncertain for the ensemble of objects, but do reflect the rank
order, from slowest to largest ejection velocity \citep{Ragozzine2007}.

The ejection velocities of 2008 AP$_{129}$, 2009 YE$_7$, and 2014 FT$_{71}$ have
been calculated by simulations
in this work. These results are based on 50 Myr-averaged orbital elements for both the
observed orbits and the orbits of test particles in simulated
clouds. We considered the nominal orbit plus
two orbits with
$3\sigma$
uncertainties in a-e space and
required the clouds of test particles to cover the three orbits in
order to determine
the minimum ejection velocity of the cloud of test particles. In the case of 2014 FT$_{71}$
the nominal orbit and one other orbit have been influenced by the 7:4 mean motion
resonance with Neptune, whereas one orbit is not influenced by this resonance
and resulted in a significantly higher ejection velocity of 178$\pm$2 m/s than our
preferred result of 30$\pm$1 m/s.

\begin{table}
\caption{Diameters and geometric albedos of probable dynamical interlopers of the Haumea family.}
\begin{tabular}{lllll}
\hline
Name                     & Diameter          & Geometric                    & Ref. & Cause of \\
                         & (km)              & albedo                       &      & exclusion \\
\hline
1996 TR$_{66}$  & \ldots            & \ldots                       &      & NIR colors \\
1999 KR$_{16}$  & 232$_{-36}^{+34}$ & 0.105$_{-0.027}^{+0.049}$    & (1)  & Very red  \\
\noalign{\smallskip}
2002 AW$_{197}$ & $768_{-38}^{+39}$ &  $0.112_{-0.011}^{+0.012}$   & (3)  & NIR spectra  \\
\noalign{\smallskip}
1999 RY$_{215}$ & $263_{-37}^{+29}$ & $0.0325_{-0.0065}^{+0.0122}$ & (3)  & (J-H$_\mathrm{S}$) color \\
\noalign{\smallskip}
Salacia         & $901 \pm 45$      & 0.044$\pm$0.004              & (4)  & NIR spectra \\
Makemake        & 1430$\pm$9        & 0.77$\pm$0.03                & (2)  & Methane ice  \\
1998 WT$_{31}$  & \ldots            & \ldots                       &      & Red slope \\
2005 UQ$_{513}$ & $498_{-75}^{+63}$ & $0.202_{-0.049}^{+0.084}$    & (3)  & Red slope \\
1996 RQ$_{20}$  & \ldots            & \ldots                       &      & Very red  \\
1999 CD$_{158}$ & $<$310            & $>$0.13                      & (1)  & Very red  \\
1999 OH$_4$     & \ldots            & \ldots                       &      & NIR colors \\
2000 CG$_{105}$ & \ldots            & \ldots                       &      & NIR colors \\
2001 FU$_{172}$ & \ldots            & \ldots                       &      & Red slope \\
2001 QC$_{298}$ & $303_{-32}^{+29}$ & $0.063_{-0.018}^{+0.029}$    & (3)  & (J-H$_\mathrm{S}$) color \\
2002 GH$_{32}$  & $<$230            & $>$\,0.13                    & (3)  & Very red  \\
2003 TH$_{58}$  & \ldots            & \ldots                       &      & (J-H$_\mathrm{S}$) color \\
2004 PT$_{107}$ & $400_{-51}^{+45}$ & $0.033_{-0.007}^{+0.011}$    & (3)  & (J-H$_\mathrm{S}$) color \\
2005 GE$_{187}$ & \ldots            & \ldots                       &      & (J-H$_\mathrm{S}$) color \\
2010 KZ$_{39}$  & \ldots            & \ldots                       &      & NIR colors \\
\hline
\end{tabular}
\tablefoot{Diameter is given only if measured by thermal radiometric techniques or by occultations.
The upper/lower limits of size/albedo are based on 2$\sigma$ flux density
limits of the most constraining wavelength band.
The reason used in the literature (e.g. \citealp{Snodgrass2010}) to reject a target as a family
member is given in the last column. \\
\\
\textbf{References.} (1) this work, (2) \citealp{Ortiz2012a}, (3) \citealp{Vilenius2014}, (4) \citealp{Fornasier2013}.}
\label{rejected}
\end{table}

\subsection{Size and albedo distributions}
\label{albedodiscussion}
\begin{figure}
   \includegraphics[width=10cm]{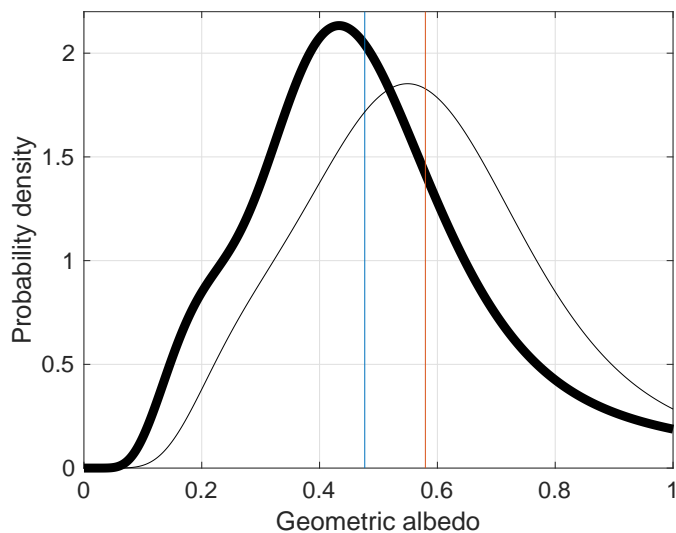}
   \caption{Combined probability density distribution of geometric albedos of
   confirmed Haumea family members 2002 TX$_{300}$, 2003 OP$_{32}$, 2005 RR$_{43}$,
   and 2003 UZ$_{117}$. The thick line is the albedo distribution assuming the
   solutions based on the beaming factor value $\eta$=1.74$\pm$0.17 for
   2003 OP$_{32}$, 2005 RR$_{43}$, and 2003 UZ$_{117}$ and the thin line
   assuming the solutions with the canonical beaming factor value
   $\eta$=1.20$\pm$0.35. The median values of the two distributions are
   $p_V=0.48_{-0.18}^{+0.28}$ (blue vertical line indicates the median) for the preferred solutions and
   $p_V=0.58_{-0.21}^{+0.27}$ (red vertical line) assuming the canonical beaming factor.}
   \label{albedodistr}
\end{figure}

We have constructed a combined probability density distribution of geometric
albedos based on the few measured targets. The asymmetric uncertainties have
been taken into account using the approach of \cite{Mommert2013}.
Instead of having two tails from a normal distribution, which would
create a discontinuity in case of asymmetric error bars, we use a log-normal
distribution.\footnote{If 63.8\% of albedo values in a normal distribution are
located within [$p_V-\sigma_{p_V}^-$,$p_V+\sigma_{p_V}^+$], where
$\sigma_{p_V}^-$ and $\sigma_{p_V}^+$ are the asymmetric uncertainties,
then the equivalent amount
is located within $[p_V/\exp\left( \sigma \right)$,$p_V \exp \left( \sigma \right)$]
in a log-normal distribution with shape parameter $\sigma$. The shape parameter
is determined by setting $p_V+\sigma_{p_V}^+$=$p_V \exp \left( \sigma \right)$
or $p_V-\sigma_{p_V}^-$=$p_V/\exp\left( \sigma \right)$; for practical implementation,
see Appendix B.2.2 in \citet{Mommert2013}.}
The combined geometric albedos (Fig.~\ref{albedodistr}) of
four Haumea family members that have measured geometric albedos
(from Table~\ref{table_results}) have a median\footnote{The error bars
of this median are calculated by finding the $p_V$ points of the c.d.f.
of geometric albedo where the value is $\frac{1-\mathrm{erf}(1/\sqrt{2})}{2}$
and $\frac{1+\mathrm{erf}(1/\sqrt{2})}{2}$, for the lower and upper
uncertainties, respectively.} of
$p_V$=0.48$_{-0.18}^{+0.28}$ using the fixed-$\eta$ solutions based on
Haumea's beaming factor for 2003 OP$_{32}$, 2005 RR$_{43}$, and 2003 UZ$_{117}$
(the geometric albedo of 2002 TX$_{300}$ is derived from a stellar occultation)
and $p_V$=0.58$_{-0.21}^{+0.27}$ if the canonical beaming factor is used instead.

\begin{figure}
   \includegraphics[width=10cm]{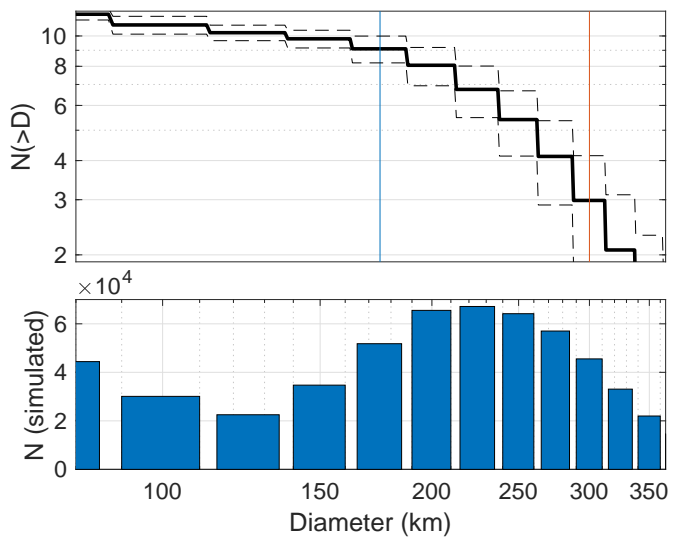}
   \caption{Combined statistical distribution of sizes (measured if available, otherwise
   inferred from the albedo distribution and $H_V$) of confirmed
   Haumea family members, including the moons. The bin size is 25 km.
   The size range 150--300 km for which the slope parameter is determined
   is indicated by the blue and red vertical lines.
   The lower panel shows
   the size histogram of 50000 randomly generated objects (see text).}
   \label{sizedistr}
\end{figure}

We have measured sizes for four confirmed family members (other than Haumea).
For the other family members absolute visual magnitudes are available. The
size distribution of the Haumea family, excluding Haumea
(Fig.~\ref{sizedistr}), is constructed in a statistical way by using measured
size values when available and otherwise
by assigning an albedo from the distribution shown in Fig.~\ref{albedodistr}
and using the absolute visual magnitudes $H_V$
(Table~\ref{table_overview_groundobs}).
Size distributions are formed 50000 times so that each measured or inferred size
may vary according to its error bar.
The slope parameter\footnote{We determine the size distribution $N(>D) \propto D^{1-q}$.}
in the size range 175-300
km is $q=$3.2$_{-0.4}^{+0.7}$.
All the measured effective diameters are
$>150$ km and the decrease of the slope below this size may be due
to an incomplete sample in the size bins $<150$ km (see the lower panel of
Fig.~\ref{sizedistr}) as only two confirmed family members
(see Table~\ref{members}) have size estimates $<100$ km based on
the assumed albedo.
If instead of using the sizes and albedo distribution based
on the fixed-$\eta$ value of 1.74 we use solutions based on the canonical value
of 1.20 (see Table~\ref{table_results} and Fig.~\ref{albedodistr}), then the slope
is steeper $q$=3.8$_{-0.5}^{+0.9}$ although it is within the uncertainties of the
preferred solution. However, sizes are generally smaller and geometric albedos
higher when the canonical beaming factor has been used and there are less
simulated objects in the 300 km size bin.
Considering the size range 150-275 km (i.e. excluding
the last size bin) gives a result that is similar to the nominal solution:
$q=$3.1$_{-0.4}^{+0.7}$.

The slope of the size distribution obtained here can be compared with the
slope of dynamically hot CKBOs since most of the family members and probable dynamical
interlopers belong to that class. The large end of the size distribution of
dynamically hot CKBOs is $q=$4.3$\pm$0.9 \citep{Vilenius2014} turning into a
shallower slope of $q=2.3\pm$0.1 in the size range 100-500 km.
We have also determined the size distribution of $<$500 km probable dynamical
interlopers from Table~\ref{rejected} (using average geometric albedo of
dynamically hot CKBOs from \citet{Vilenius2014} and $H_V$ from MPC when no measured size
available): q=2.0$\pm$0.6, which is compatible with the slope parameter of
the general hot CKBO population. Comparing the two above-mentioned slope
parameters to those determined for the Haumea family ($q\sim3$) indicates that
the family has a slope that is steeper than the background population of
dynamically hot CKBOs in the same size range.

There are different models for the slope of the size distributions of
collisional fragments in the literature. The value determined in this work is approximately
compatible with the classical slope of -2.5 \citep{Dohnanyi1969,Carry2012},
which corresponds to $q=3.5$ in our definition of the slope parameter.

\subsection{Albedo and family membership}
The albedos and diameters of the TNOs assumed to be dynamical
interlopers in the Haumea family are given in Table~\ref{rejected}.
The table
also briefly summarizes the rationale for excluding each object
from inclusion as a true member of the collisional family. The
albedo values are an independent data set that bears on the
question of family membership. Excluding Makemake and Salacia, each
of comparable size to Haumea and therefore inconsistent with the
assumption that Haumea itself defines the centre of mass for the
collisional family, the median geometric albedo for these objects
is 0.08$_{-0.05}^{+0.07}$. Most of the objects in Table~\ref{rejected}
are dynamically hot CKBOs,
and their average albedo is consistent with the average albedo of
those objects, $p_V$ = 0.085$_{-0.045}^{+0.084}$ \citep{Vilenius2014}. As
discussed in Sect.~\ref{albedodiscussion}, the average albedo
measured for the four accepted family members is 0.48, much higher than for the
objects in Table~\ref{rejected}.
This suggests that these objects do not have albedos similar to
those of the accepted family members (although the sample sizes, six
interlopers and four family members, are very small). Of the objects
in Table~\ref{rejected}, only the hot CKBO 2005 UQ$_{513}$ has an unusually high
albedo, and its albedo is significantly higher than the average for
hot CKBOs in general. Our results for
1999 CD$_{158}$ and 2002 GH$_{32}$ suggest that they may also have high
albedos. Table~\ref{rejected} gives the 2$\sigma$ lower limits on albedo and
upper limits on size (i.e. the probability that the geometric albedo is
$<$0.13 is $\approx$4.6\%). In summary, albedo measurements for
objects previously identified as dynamical interlopers seem to
support that identification in general, but suggest that three
of them may have unusually high albedos, and further investigation
may be warranted. It is unfortunate that there is not more data
constraining the albedos and diameters of both the interlopers
and the family members.

The Haumea family members have
significantly higher albedos than the averages of scattered disk,
detached, or cold CKBOs, which are the dynamical classes with
the highest average albedos \citep{SantosSanz2012,Vilenius2014}.
For mid-sized TNOs, such as Haumea family members,
the high albedo surface indicates lack of hydrocarbons, which would have
produced a darker and redder surface over long periods of exposure to space
weathering~\citep{Brown2012}.
This is compatible with the collisional hypothesis,
which states that the fragments are high-albedo water ice
pieces from the mantle of proto-Haumea.
In a colour-albedo plot the Haumea family members, with their high
albedos, are distinct from the probable dynamical interlopers, which are more
widely spread in the colour-albedo plot of \citet[Fig. 2]{Lacerda2014b}.

\subsection{Mass and ejection velocity}
The masses of Haumea's moons Hi'iaka and Namaka are
(20.0$\pm$1.2)$\times 10^{18}$ kg and (2.0$\pm$1.6)$\times 10^{18}$ kg
\citep{Ragozzine2009,Cuk2013}.
For the other family members we estimate masses assuming bulk densities of
1 g cm$^{-3}$. The confirmed members would constitute approximately 2.4\% of
the mass of Haumea (using sizes from Table~\ref{members} when no mass or size
measurement available). The largest family member, the moon Hi'iaka, would alone
constitute 21\% of the mass of the family excluding Haumea and the five largest
family members would
be more than half of the total mass of the family (excluding Haumea). Using
the alternative radiometric solutions of Table~\ref{table_results} and the
lower median geometric albedo results in a mass estimate of 2.0\%. If all the
candidate family members in Table~\ref{potentials} were be confirmed, they
would constitute $\sim$0.2\% of Haumea's mass.

The scenarios in which the proto-moon of Haumea underwent fission to produce a family presented
by \cite{Ortiz2012b} require that the mass of the moons and the family members
is less than 20\% of Haumea's mass. \cite{Ortiz2012b} had only one measured albedo
available (2002 TX$_{300}$) and they used a default geometric albedo of 0.6
for other family members. Our new observations give more confirmation in
using a high albedo and our new mass estimate of the family
is compatible with the mass ratio assumption used by \cite{Ortiz2012b}.
Our mass estimate of 2.4\% does not exclude the formation mechanisms by
disruption of a large satellite of proto-Haumea \citep{Schlichting2009}, which
predicts an upper mass ratio limit of 5\%. The mechanism proposed by \citet{Leinhardt2010},
where two equal-sized objects merge, predicts a mass ratio of 4-7\%
\citep{Volk2012}, which is higher than our current estimate.

\begin{figure}
   \includegraphics[width=10cm]{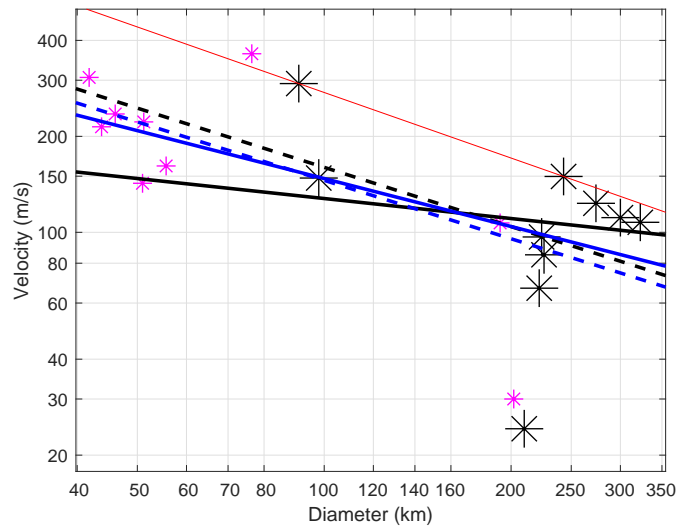}
   \caption{Distribution of ejection velocities. The large black dots are
   confirmed Haumea family members (Table~\ref{members}) and the magenta dots
   are candidate family members (Table~\ref{potentials}). The dashed black line is
   a fit to all confirmed family members ($\gamma$=0.62);
   the solid black line is a fit to the same data but excluding the minimum
   and maximum velocities ($\gamma$=0.21).
   The red line is a limiting case of the confirmed family members
   ($\gamma=0.68$).
   The dashed blue line is a fit to all confirmed and candidate family members ($\gamma$=0.61)
   and the solid blue line a fit to the same data excluding
   minimum and maximum velocities ($\gamma=0.50$).}
   \label{velodistr}
\end{figure}

The ejection velocity of a fragment and its mass are related via a power law.
Assuming a constant density for all family members, this relation may be
written in terms of the diameter as \citep{Lykawka2012}
$v_e \propto D^{-\gamma}$,
where $v_e$ is the ejection velocity and the power-law slope
$\gamma$ is $\lesssim$
0.5 \citep{Zappala2002}. Figure~\ref{velodistr} shows a fit to effective diameters
and velocities from Table~\ref{members}. A fit using confirmed family members
gives $\gamma$=0.62, which is slightly larger than
the upper limit of plausible values.
\cite{Lykawka2012} and references cited therein note that there is often large
scatter in the ejection velocity values and large ratios of maximum-to-minimum
values. Therefore, another fit is made by ignoring the minimum and maximum
velocities (1996 TO$_{66}$ and 1999 OY$_3$). This gives a lower value of
$\gamma$=0.21. We have repeated the fit with an
extended data set including all the candidate family members
(Table~\ref{potentials} and inferred sizes using the geometric albedo
distribution of the Haumea family). The extended data set gives $\gamma$=0.61
for all data and $\gamma$=0.50
when minimum and maximum velocities are excluded (1996 TO$_{66}$ and 2003 HX$_{56}$).
In the above calculations we used absolute visual magnitudes and an assumed
geometric albedo
of $p_V$=0.48 to assign diameters to objects lacking a measured size. If the
canonical value of the beaming factor is used in fixed-$\eta$ solutions of the
family members, the resulting median geometric albedo is higher: $p_V$=0.58. With this
geometric albedo the result (confirmed and candidate family members excluding
minimum and maximum velocities) is $\gamma$=0.46, which indicates that the result
is not sensitive to a moderate difference in the assumed geometric albedo.

The fitted values indicate that ejection velocities
are dependent on diameter although in some of the cases the power-law slope is
0.5-0.6, which is higher than expected from theory. This means that smaller
fragments of the
Haumea family have been dispersed in the orbital element space much more than
the currently known larger fragments \citep{Lykawka2012}. This may affect
theories of the formation of the family that try to solve
the problem of too low velocities: an average based on those velocities is
probably biased by the fact that we have only observed, and discovered,
the largest fragments of the family, which have lower velocities than smaller
fragments.

\subsection{Correlations}
The small number of reliably measured Haumea family members and dynamical interlopers
makes it challenging to detect correlations. The diameter and geometric albedo
results (Table~\ref{table_results}) suggest a positive
correlation but, when taking
into account the error bars \citep{Peixinho2015}, for the Haumea family objects we obtain a
Spearman correlation coefficient of $\rho_{\,D\,p_V}=0.65 ^{+ 0.22 }_{ -0.45 }$
with a P-value of 0.40 (i.e. confidence level CL$=0.84\,\sigma$), being, therefore, not significant.
For the eight dynamical interlopers (Table~\ref{rejected}) the correlation strength appears weaker,
but the evidence is of the same order as in the Haumea family members and also not significant:
$\rho_{\,D\,p_V}=0.37 ^{+ 0.38 }_{ -0.54 }$ (P-value$=0.38$; CL$=0.88\,\sigma$).

Nevertheless, some remarks about minimum sampling for detections can be made.
Supposing that the correlation between effective diameters and geometric albedos among
the Haumea family was $\rho=0.9$, given our error bars and the low dispersion
of albedos and diameters, such a correlation
would be observationally ``degraded'' to $\sim 0.65$ \citep[see][]{Peixinho2015}
and we would need a sample of $n \geq 39 $ objects to have a risk
lower than $10\%$ of missing it, if we aim at a
$3\sigma$
level detection.
Analogously, regarding the dynamical interlopers, even if their true diameter-albedo correlation
was $\rho=0.4$, we would need a sample of $n \geq 112$ objects to ensure the detection.
Most of the dynamical interlopers are classified as dynamically hot CKBOs
and a sample of 26 objects in that class (excluding Haumea family and dwarf planets)
showed no evidence of a diameter-albedo correlation at 3$\sigma$ level taking
into account the error bars \citep{Vilenius2014}.

To confirm that the diameter-albedo correlation
among the Haumea family objects would indeed be different from the one among the dynamical
interlopers, at a $3\sigma$
level, we would need to increase the sampling required
to detect the presence of the correlations by a factor of $2.5$ compared to
the numbers of objects given above.
 The accuracy of size and
albedo estimates can improve in the future, for example by more stellar
occultations. If the error bars were lower than $\sim 5\%$, then a sample of 15
Haumea family objects and 112 dynamical interlopers would be enough to confirm
a difference between $\rho$=0.9 and $\rho$=0.4 at a 3$\sigma$ level.

\section{Conclusions}
We have measured the sizes and geometric albedos of three confirmed Haumea
family members: 2003 OP$_{32}$, 2005 RR$_{43}$, and 2003 UZ$_{117}$.
In addition, we have updated the results of 2002 TX$_{300}$, 1996 TO$_{66}$,
1995 SM$_{55}$, and 1999 KR$_{16}$.
We have also refined or determined optical phase coefficients for several
family members and candidate members and have determined the ejection
velocities of 2008 AP$_{129}$, 2009 YE$_{7}$, and 2014 FT$_{71}$.
The ejection velocity is inversely correlated with the fragment diameter, and
therefore the Haumea family may be less compact than thought.
An average ejection velocity is probably biased by the fact that we have only
observed, and discovered, the largest fragments of the family, which have lower
velocities than smaller fragments.

Our analysis has utilized
the results of the stellar occultation by Haumea \citep{Ortiz2017} and has the
following main conclusions:

\begin{enumerate}[-]
\item Our measurements indicate
that
Haumea family members have a diversity of high to very high albedos
and the lowest albedo among the detected objects is $\sim$0.29
and the albedo limit of non-detected targets is $\sog$0.2,
which is higher than the average albedo of TNOs ($\sim$0.10). The median albedo
of the Haumea family is $p_V$=0.48$_{-0.18}^{+0.28}$. The highest-albedo
member is 2002 TX$_{300}$.
\item The median geometric albedo of probable dynamical interlopers in the
Haumea family is 0.08$_{-0.05}^{+0.07}$, consistent with that of the dynamically hot
CKBO population, and much lower than that for the accepted family members.
Object 2005 UQ$_{513}$ does have an unusually high albedo (0.22), and two other
objects (1999 CD$_{158}$ and 2002 GH$_{32}$) have 2$\sigma$ lower limits on their
albedos of 0.13. Many Haumea family members and dynamical relatives lack albedo
determinations, making interpretation of these albedo results tentative, but
there is no strong evidence based on albedo that any of the dynamical
interlopers should be considered as possible family members.
\item Using measured sizes when available and an average albedo with optical absolute
brightness for other family members, we determine the cumulative size
distribution and find its slope to be $q$=3.2$_{-0.4}^{+0.7}$ for diameters 175$<$D$<$300 km.
This is steeper than the slope of dynamically hot CKBOs in general in the same size range.
\item We estimate the confirmed family members and the two moons to constitute
2.4\% of the mass of Haumea.
\item The ejection velocity depends on diameters of the fragments with a power-law slope of 0.21
(ignoring the minimum and maximum velocities). If candidate
family members are included, to cover a broader diameter range, the slope is steeper: 0.50.
\item We have determined Haumea's beaming factor: $\eta$$=$$1.74_{-0.17}^{+0.18}$,
which indicates a thermal inertia of $\Gamma$$\sim$1 $\mathrm{J} \mathrm{m}^{-2} \mathrm{K}^{-1}
\mathrm{s}^{-\frac{1}{2}}$.
\end{enumerate}

\label{conclude}

\begin{acknowledgements}
Part of this work was supported by the German \emph{DLR} project number 50 OR
1108.
TM, CK, PS, and RD acknowledge that the research leading to these results
   has received funding from the European Union's Horizon 2020
   Research and Innovation Programme, under Grant Agreement no
   687378.
AP acknowledges the grant LP2012-31 of the Hungarian Academy of Sciences.
NP acknowledges funding by the Portuguese FCT - Foundation for Science and Technology (ref: SFRH/BGCT/113686/2015).
CITEUC is funded by Portuguese National Funds through FCT - Foundation for Science and Technology
(project: UID/ Multi/00611/2013) and FEDER - European Regional Development Fund through
COMPETE 2020 - Operational Programme Competitiveness and Internationalisation
(project: POCI-01-0145-FEDER-006922).
C.K. has been supported by the K-125015 and GINOP-2.3.2-15-2016-00003 grants of the National Research, Development and Innovation Office (NKFIH,
Hungary).

\end{acknowledgements}

\bibliographystyle{aa}
\bibliography{Vol618A136_2018}


\end{document}